\documentclass[12pt,preprint]{aastex}

\shorttitle{The Search for Muon Neutrinos from GRBs with AMANDA}
\shortauthors{The IceCube Collaboration}

\begin{document}

\title{The Search for Muon Neutrinos from Northern Hemisphere Gamma-Ray Bursts with AMANDA}

\author{
A.~Achterberg\altaffilmark{1},
M.~Ackermann\altaffilmark{2},
J.~Adams\altaffilmark{3},
J.~Ahrens\altaffilmark{4},
K.~Andeen\altaffilmark{5},
J.~Auffenberg\altaffilmark{14},
J.~N.~Bahcall\altaffilmark{7,34},
X.~Bai\altaffilmark{8},
B.~Baret\altaffilmark{9},
S.~W.~Barwick\altaffilmark{11}
% ...
}

\altaffiltext{1}{Dept.~of Physics and Astronomy, Utrecht University/SRON,
  NL-3584 CC Utrecht, The Netherlands}
\altaffiltext{2}{DESY, D-15735 Zeuthen, Germany}
\altaffiltext{3}{Dept.~of Physics and Astronomy, University of Canterbury,
  Private Bag 4800, Christchurch, New Zealand}
\altaffiltext{4}{Institute of Physics, University of Mainz, Staudinger
Weg 7,
  D-55099 Mainz, Germany}
\altaffiltext{5}{Dept.~of Physics, University of Wisconsin, Madison, WI
53706,
  USA}
% ...
\altaffiltext{7}{Institute for Advanced Study, Princeton, NJ 08540, USA}
\altaffiltext{8}{Bartol Research Institute, University of Delaware,
Newark, DE
  19716, USA}
\altaffiltext{9}{Vrije Universiteit Brussel, Dienst ELEM, B-1050 Brussels,
  Belgium}
% ...
\altaffiltext{11}{Dept.~of Physics and Astronomy, University of California,
  Irvine, CA 92697, USA}
% ...
\altaffiltext{14}{Dept.~of Physics, University of Wuppertal, D-42119
  Wuppertal, Germany}
% ...
\altaffiltext{34}{Deceased}
% ...

\email{* Corresponding author K. Kuehn: kuehn@HEP.ps.uci.edu}


\begin{abstract}
We present the results of the analysis of neutrino observations by the Antarctic Muon and 
Neutrino Detector Array (AMANDA) correlated with photon observations of more than 400 
gamma-ray bursts (GRBs) in the Northern Hemisphere from 1997 to 2003.  During this 
time period, AMANDA's effective collection area for muon neutrinos was larger than that of 
any other existing detector.  Based on our observations of zero neutrinos during and 
immediately prior to the GRBs in the dataset, we set the most stringent upper limit on muon 
neutrino emission correlated with gamma-ray bursts.  Assuming a Waxman-Bahcall spectrum and 
incorporating all systematic uncertainties, our flux upper limit has a normalization at 1 
PeV of

\noindent
$E^{2}$${\Phi}_{\nu}$ ${\leq}$ 6.0 $\times$ $10^{-9}$ $\mathrm{GeV cm^{-2} s^{-1} sr^{-1}}$,

\noindent
with 90\% of the events expected within the energy range of $\sim$10 TeV to $\sim$3 PeV.  
The impact of this limit on several theoretical models of GRBs is discussed, as well as the 
future potential for detection of GRBs by next generation neutrino telescopes.  Finally, we 
briefly describe several modifications to this analysis in order to apply it to other types 
of transient point sources.
\end{abstract}


\section{Introduction}

\subsection{Gamma-Ray Bursts}

Gamma-ray bursts (GRBs) are among the most energetic phenomena in the universe; based on 
their luminosity and the cosmological distances derived from redshift measurements of burst 
afterglows and/or host galaxies\citep{Beppo}, GRBs require the release of an enormous amount of 
energy (E $\approx$10$^{53}$ $\times$ $\Omega$/4$\pi$ erg, where $\Omega$ is the solid angle of 
the GRB jet) in as little as a fraction of a second\citep{Frail}.  Based on the observations of 
the Burst and Transient Source Experiment\citep[BATSE, see][]{BATSE} and other space-based 
detectors, they are expected to occur throughout the observable universe at a rate of 
${\gtrsim}$700 per year, though current instrument do not have sufficient sky coverage or 
sensitivity to detect every burst.  Long duration (${\gtrsim}$2 sec) bursts are believed to 
originate from the collapse of a massive stellar progenitor into a black hole, whereas 
short duration (${\lesssim}$2 sec) bursts are believed to result from the merger of two 
compact objects into a black hole \citep{Eicher}\footnote{For a more recent treatment of 
the compact object merger scenario, see \citep{Pacz, Lewin}, and for an alternative 
description of the GRB progenitor scenario, see also \citep{Roming}.}.  Though these two 
types of bursts come from different progenitors, both are consistent with the canonical 
picture of gamma-ray bursts---the fireball scenario \citep{Fryer,Piran1}.  A fireball is 
generated during the formation of the black hole when the outflowing plasma is accelerated 
to ultrarelativistic speeds.  Subsequently, in an optically thin region (outside of the 
progenitor), the kinetic energy of the plasma is converted to radiation, either through 
interaction with an external medium or through self-interaction within the flow 
\citep{Piran2}.  If the circumstellar environment contains enough baryonic material, it 
will be entrained with the accelerated plasma.  Subsequent photo-pion production by baryon 
interaction with synchrotron or inverse Compton scattered photons will lead to several 
decay products, including neutrinos and antineutrinos in a ratio of 2:1.  The primary 
reaction is:
\begin{equation}
p + \gamma \rightarrow \Delta \rightarrow \pi^{+} + n
\end{equation}
followed by 
\begin{equation}
\pi^{+} \rightarrow \mu^{+} + \nu_{\mu} 
\end{equation}
after which the muon will further decay to
\begin{equation}
\mu^{+} \rightarrow e^{+} + \nu_{e} + \bar{\nu_{\mu}}.
\end{equation}
\noindent
Similarly, ``precursor'' neutrinos may be generated by p-p interactions either within the 
star or in the immediate circumburst environment (see Section 2).  Due to their minuscule 
interaction cross section \citep{Gandhi}, neutrinos will reach the AMANDA detector after 
traveling nearly unimpeded from the GRB environment.  AMANDA has been searching for 
high-energy neutrinos from various astrophysical fluxes (both discrete and diffuse) for 
nearly a decade; in this work we focus on the analysis of AMANDA data correlated with 
photon observations of more than 400 GRBs from 1997 to 2003.

\subsection{The AMANDA Detector}

The AMANDA detector \citep{AMANDA} is an array of Optical Modules (OMs) deployed at depths 
between 1.5 and 2 km beneath the surface of the ice at the South Pole.  An OM consists of a 
photomultiplier tube housed in a glass pressure sphere. During the years 1997-1999 the 
detector operated with 302 OMs on ten strings placed in a circular geometry with a diameter 
of about 100 m, and was known as AMANDA B-10.  From 2000 onward, nine additional strings 
were in operation, placed within a diameter of about 200 m, bringing the total number 
of optical modules to 677.  This phase of the neutrino observatory (dubbed AMANDA-II) 
operated through 2004, and continues as a high density component of IceCube, a km-scale 
detector currently being constructed \citep{IceCub}.

The optical modules in AMANDA are designed to detect the Cherenkov emission from 
neutrino-induced muons that travel through or near the instrumented volume of ice.  While 
other neutrinos may be detected with this search, the efficiency for ${\nu}_{e}$ or 
${\nu}_{\tau}$ detection is significantly smaller.  Other multi-flavor GRB neutrino searches 
which don't require directional information have been performed \citep{Casc}; we focus 
here on the search for GRB muon neutrinos from the Northern Hemisphere ($\delta$ from 0$^{\circ}$ to 90$^{\circ}$).  Due to the limited 
volume of ice above the detector, few downgoing extraterrestrial neutrinos will interact 
above and be detected by AMANDA.  At the energies of interest to this analysis, the 
down-going events in the AMANDA dataset are primarily the atmospheric muon background which 
will completely overwhelm any potential downgoing signal.  Thus, our extraterrestrial 
signal is primarily confined to the horizontal or up-going direction.  As these muon 
neutrinos travel through the ice, they may interact with nearby nucleons to create 
energetic muons:
\begin{equation}
\nu_{\mu} + N \rightarrow \mu + X,
\end{equation}

\noindent
where N is a nucleon and X represents other reaction products.  Muons produced in this 
reaction can carry a significant fraction of the original neutrino energy
\citep{Gandhi2}.  Depending on its energy, the muon can travel up to tens 
of kilometers through the ice; for $\nu_{\mathrm{\mu}}$ in the energy range of greatest 
interest to AMANDA (${\sim}10^{5}$ GeV), the muon path length is ${\sim}$10 
km \citep{Lipari}.

Since AMANDA can detect such a muon anywhere along its substantial path length, the 
effective detector volume is significantly larger than the actual instrumented volume.  A 
muon that has sufficient energy will continuously emit Cherenkov radiation, and will also 
generate additional particles due to stochastic processes.  The ice at a depth of more than 
one kilometer is extremely clear, and thus the Cherenkov photons have large scattering 
($L^{\mathrm{{eff}}}_{\mathrm{s}}$) and absorption ($L_{\mathrm{a}}$) lengths---at 
$\lambda$ = 400 nm, $L^{\mathrm{eff}}_{\mathrm{s}}$${\approx}$25 m and 
$L_{\mathrm{a}}$${\approx}$100 m \citep{Kurt}.  The Cherenkov light therefore has the 
potential to reach numerous OMs as the muon travels through the detector, and the relative 
timing of the hit OMs provides the basis for a set of maximum-likelihood reconstruction 
algorithms to determine the muon's direction of origin \citep{Reco}.  The algorithms 
applied to this analysis are based on variations from a randomly-seeded ``first guess'' 
track using the Pandel function to parametrize the sequence of OM hits.  The likelihood of 
the initial track is calculated, and then the procedure is iterated (up to 32 times) to 
determine the most likely muon track.  Iterations beyond the first incorporate increasingly 
complex features of the detector response to the Cherenkov photons, the details of which 
are beyond the scope of this work\footnote{An alternative track reconstruction known as a 
``paraboloid fit'' is also relevant for our secondary data selection criteria, see 
Section 3.3 for further details.}.  Detector simulations, along with observations of 
downgoing cosmic ray muons, have shown that this procedure provides track reconstructions 
accurate to within a mean value of ${\sim}$2$^{\circ}$.  Atmospheric muons are almost 
entirely removed from the dataset by constraining our search to those bursts occurring in 
the Northern Hemisphere, allowing the detector to be shielded from a substantial background 
flux by the bulk of the earth.  Upgoing atmospheric neutrinos caused by cosmic ray 
interactions in the Northern hemisphere may also be detected by AMANDA, as their spectrum 
extends into the energy range of relevance to the GRB search.  However, they likewise are 
removed from the dataset by requiring strict spatial and temporal correlation with photon 
observations of GRBs.  With these selection criteria applied, we expect less than 0.01 
atmospheric neutrino events in our dataset.

In Section 2 we describe several models for GRB neutrino emission.  In Section 3 we discuss 
the method for determining periods of stable detector performance and for separating the 
expected GRB neutrino signal from all misreconstructed background events, as well 
as the systematic uncertainties associated with this analysis procedure.  In Section 4 we 
compare the results of the AMANDA observations with the models, as well as provide a 
spectrum-independent method for determining the fluence upper limit from GRBs.  We conclude 
with the future potential of AMANDA/IceCube, for both the standard GRB search in the Swift 
era \citep{Mark} and for searches optimized for other transient point sources, such as 
jet-driven supernovae.

\section{Models of Neutrino Emission}

According to the canonical description provided above, gamma-ray bursts result from the 
dissipation of the energy of relativistic outflows from a central engine.
Based on the assumption that GRBs are the source of ultra-high energy cosmic rays (UHECRs), 
Waxman and Bahcall predicted an annual muon neutrino flux associated with GRBs of 
$E^{2}$$\Phi_{\mathrm{\nu}}$ $\sim$ 9 $\times$ $10^{-9}$ $\mathrm{GeV cm^{-2} 
s^{-1} sr^{-1}}$ from 100 TeV to 10 PeV \citep{Waxman} 
\footnote{For the original formulation of this neutrino flux prediction, see 
\citep{WaxBah}.  Note that this GRB neutrino flux is distinct from the Waxman-Bahcall upper 
bound on the diffuse neutrino flux due to UHECRs.}.  \citet{Murase} predict a similar 
spectrum to Waxman and Bahcall for long-duration bursts, though their simulations include a 
wider range of parameters, leading to a wider variation in predicted fluxes.  Inclusion of 
neutrino oscillations reduce these predictions by a factor of two\footnote{Oscillations 
modify the flavor ratio from 1:2:0 at the source to 1:1:1 at Earth.  However, see 
\citep{Kashti} for a discussion regarding different flavor ratios due to energy losses 
of the ${\pi}$ and ${\nu}$.}.  \citet{Razzaq} hypothesize a different scenario in which a 
supernova precedes a long-duration GRB by several days to a week.  In this ``supranova'' 
scenario, the supernova remnant provides target nucleons for pp interactions leading to 
precursor neutrinos with energy $E_{\nu}$${\sim}$10 TeV. Furthermore, the remnant will 
produce target photons for p${\gamma}$ interactions, which will also yield muon neutrinos 
up to $10^{16}$ eV, albeit with a different spectral shape than that predicted by the 
Waxman-Bahcall model\footnote{Though the supranova model is still within the realm of 
possibility, it is somewhat disfavored based on observations of GRB060218, in which the 
supernova preceded the GRB by at most a few hours---not long enough to provide an ideal 
circumburst environment for a significant neutrino flux.}.  This model also has 
implications for gamma-ray dark (or ``choked'') bursts, which are briefly discussed in 
Section 5.  Furthermore, though these models explicitly incorporate only long-duration 
bursts into their models, the GRB central engine is in principle independent of burst type.  
Thus, though the flux upper limits for these models include long bursts only, the models 
could potentially be expanded to include neutrinos from short bursts as well.  Within the 
AMANDA dataset long bursts dominate over short bursts; incorporating short bursts would 
have a small, though not insignificant, effect on the overall limit (see Section 4 for 
details).  Figure 1 shows the expected GRB neutrino flux based on four representative 
models.  The precursor model predicts a neutrino flux as early as several tens of seconds 
prior to the observed GRB photons, whereas the other models tested here predict a neutrino 
flux in coincidence with the GRB photons\footnote{Any time delay in observing the neutrinos 
due to the neutrino mass is assumed to be negligible compared to the time scale over which 
we search for the burst emission.}.  Other models of GRB emission also exist 
\citep[see, e.g.][]{Dermer1,Dermer2}; though we do not explicitly focus on such models 
here, a flux upper limit can be calculated for such models using the Green's Function 
method detailed in Section 4.

Many theoretical models (most notably, the Waxman-Bahcall model) are based on assumptions 
regarding the circumburst environment as well as the average properties of bursts (total 
emission energy, redshift, etc.) which do not correspond directly to the properties of 
specific bursts.  It is possible to estimate the muon neutrino flux for individual bursts, 
but these estimates vary substantially, and often bracket the predictions of the averaged 
properties \citep{Mike}.  For those bursts where redshift and spectral information is 
available, more accurate estimates of muon neutrino flux can be made on a burst-by-burst 
basis.  For extremely bright, nearby bursts (e.g. GRB030329), the predicted fluxes can be 
as much as two orders of magnitude greater than the mean burst flux \citep{MikeSwift}.  
Finally, our simulations assume a $\mathrm{\Phi_{\nu}}$:$\mathrm{\Phi_{\bar{\nu}}}$ ratio of 
1:1.  AMANDA does not distinguish the muon charge; however, neutrino event rates are 
larger than anti-neutrino rates for an equal flux, since the neutrino cross section is 
larger up to energies of ${\sim}$$10^5$ GeV.  Thus, any models proposing a ratio other than 
unity will result in a different expected event rate and, ultimately, a different flux upper 
limit for this analysis.  

\section{Observation Procedure}

\subsection{Correlated Observations}

This AMANDA GRB search relies on spatial and temporal correlations with photon observations 
of other instruments including BATSE aboard the Compton Gamma-Ray Observatory (CGRO), as 
well as HETE-II, Ulysses, and other satellites of the Third Interplanetary Network (IPN) 
\citep{Hurley}.  As stated previously, our search is restricted to that half of the bursts 
occurring in the Northern Hemisphere.  Furthermore, because engineering and maintenance 
work is performed on the AMANDA detector during the austral summer (December-February), 
only a few bursts from these months can potentially be observed each year.  For each GRB in 
the dataset, we search for muon neutrino emission during the coincident phase of burst 
emission.  The coincident phase is determined by either the $\mathrm{T_{90}}$ start and end 
times of the burst, or the entire duration of emission in excess of the background rate 
(for bursts without well-defined $\mathrm{T_{90}}$). A period of time before and after each 
burst is added to the search in order to accommodate the timing errors of the photon 
observations (which vary from burst to burst).  Most bursts have prompt phases lasting from 
a few seconds up to to a few tens of seconds, though there are some exceptional bursts 
lasting hundreds of seconds.  To investigate different model predictions for the bursts 
occurring during 2001-2003, we also performed an extended search for precursor neutrinos 
from 110 seconds before the burst start time until the beginning of the coincident search 
window.  BATSE observations were the sole source of data for the AMANDA B-10 analysis for 
1997--1999.  Other IPN-detected bursts were included beginning with the AMANDA-II dataset 
in 2000, and the analysis then relied exclusively on IPN data from other satellites once 
CGRO was decommissioned in May, 2000.  Additional bursts were also discovered in the BATSE 
archival data \citep{Kommers,Stern}; the relevant time periods of the AMANDA data were 
searched for muon neutrinos from these bursts as well.  We do not, however, include this 
particular subset of bursts in the flux or fluence upper limits for the models addressed in 
this work, because non-triggered bursts were not incorporated into the primary models of 
GRB neutrino emission.   The instruments participating in the Interplanetary Network 
through 2003 are given in Table 1 and the number of bursts searched in each year of AMANDA 
observations is listed in Table 2; information on the specific bursts included in this 
analysis is also available \citep{myurl}.

\subsection{Background and Detector Stability}

To determine the background rate and to establish data selection criteria for each burst, a 
larger period of one hour and 50 minutes of data is analyzed---from one hour before the 
burst to one hour after the burst, with the 10 minute period during and immediately 
surrounding the burst excluded to ensure that the data quality cuts are not determined in a 
biased fashion (a ``blind'' analysis).  Prior to determination of the data selection 
criteria, we study detector stability in this background period.  The specific stability 
criteria for AMANDA B-10 have been discussed previously \citep{Rellen}; here we describe 
the AMANDA-II stability criteria in more detail.

We perform two tests to identify non-statistical fluctuations in the data rate that could 
produce fake events (``false positives'') or unanticipated dead time (``false negatives'') 
in the detector.  The first test compares the observed event count per 10 second time 
bin to the expected, temporally uncorrelated, distribution of background events.  
This tests for any non-statistical fluctuations in data rate due to temporary instability 
in the detector.  Without this test, an upward fluctuation in the data rate not caused by 
neutrinos could potentially be misinterpreted as a signal event.  This test has three 
successive steps based on the P-value of the event rate distribution. The P-value of a data 
segment is defined as the percent difference between the RMS variation of the data event 
rate and the width of a Gaussian fit to the data rate distribution.  The first step 
identifies all those bursts with stable periods---those having a P-value of less than 6$\%$ 
(corresponding to variations of less than 1${\sigma}$ relative to the overall distribution 
of P-values).  The second step identifies bursts with marginally stable detector 
performance: 6${\%}$${\leq}$P${\leq}$12$\%$ (1--2${\sigma}$).  Additional tests are 
performed on these bursts; specifically, the data rate of the previously blinded 10-minute 
period is explored in a region of the sky far away from the GRB (the ``on-time, 
off-source'' region).  This maintains the blindness of the analysis, while allowing a more 
detailed exploration of detector stability.  Marginally stable burst periods are included 
in the analysis if they are also marginally stable in the on-time, off-source region 
(P-value less than 12$\%$), and if the event rate has only small (${\leq}3{\sigma}$) 
variations throughout the on-time, off-source region.  The vast majority of all burst time 
periods were stable according to these criteria.  The final step of this test is applied if 
the first two steps are inconclusive.  It requires any event rate variations greater than 
3$\sigma$ to occur at a significant distance from the burst time.  Two bursts fall into 
this category; they had marginally stable off-time periods and insufficient statistics for 
an on-time/off-source stability test.  However they were included in this analysis because 
the largest event rate variations were separated in time from the burst by several minutes.  
Only one time period associated with a burst in the AMANDA dataset had off-time and 
on-time/off-source P-values greater than 12\%, and this burst was excluded from the 
analysis.  Figure 2 shows the data rate per 10 seconds for a sample GRB period, overlaid 
with the Gaussian fit.  They are in very good agreement, showing a stable data rate for 
this period of detector activity.

The second test utilizes the time between subsequent events ($\mathrm{{\delta}t}$) to 
ensure that there is not an anomalously large amount of time between detector triggers.  
The amount of time between triggers can vary widely, but larger gaps occur with much less 
frequency than shorter gaps.  There is also unavoidable (but quantifiable) dead time 
between each trigger while the detector is being read out.  The overall effect of the 
expected dead time is to reduce the detector's signal acceptance by approximately 17\%, and 
this quantity has been incorporated into the expected neutrino observation rate for this 
analysis.  However, large unexpected gaps between triggers would indicate a period of 
unstable detector performance, and would mean that an otherwise detectable neutrino signal 
might not be observed during such a period.  We test the 1 hour and 50 minute time periods 
surrounding each burst to ensure that no such gaps occur.  An example of the temporal 
distribution of triggers compared with an exponentially decreasing fit to the 
$\mathrm{{\delta}t}$ distribution is shown in Figure 3.  The variations observed in the 
data for this time period are within 2${\sigma}$ of the observed fit for all values of 
$\mathrm{{\delta}t}$.  Thus there are no unexpected variations in the time between detector 
triggers, and we confirm that AMANDA is collecting data as expected occurring during the 
on-time window for this burst.  All data periods associated with GRBs that pass the first 
test also pass this second test for stable detector operation.

\subsection{Data Selection Criteria}

For those bursts determined to be stable by the above criteria, data quality cuts are then 
selected to separate the predicted signal from the observed background events.  This 
process relies primarily on the simulated signal events and the observed background events.
The simulation of the detector response to signal and background events is described in 
Ahrens et al. \citep{Reco}.  The simulation procedure uses the neutrino generation 
program NuSim \citep{Hill97} for signal event simulations.  Background events are simulated 
with CORSIKA \citep{Cors}, which implements the 2001 version of the QGSJET model of 
hadronic interactions \citep{Kalmykov}.  Once the neutrino or other cosmic-ray primaries 
are generated and propagated to their interaction vertex, we simulate the secondary 
propagation with the Muon Monte Carlo (MMC) package \citep{Dima}.  Finally, we simulate the 
AMANDA detector response with the software package AMASIM.  We then are able to compare 
simulated signal, simulated background, and observed background data.

In the case of the GRB search, the background rate is measured using the off-time window, 
where no signal is expected.  Thus, unlike other AMANDA analyses \citep{Diffuse,Point}, the 
background events do not need to be simulated, nor do the data events need to be scrambled 
in time or azimuth to retain a blind analysis procedure.  Exploring the variations between 
observed background events and simulated events does, however, ensure that we understand 
the systematic errors associated with the simulation process.  For example, Figure 4 shows 
excellent agreement in ${\mathrm{{\pounds}_{reco}}}$, the log(Likelihood) of the 
reconstructed tracks of the simulated and observed background events.  Given this level of 
agreement, the errors arising from discrepancies between the simulated and observed events 
are expected to be small.  Additionally, atmospheric neutrinos have previously been 
observed by AMANDA up to TeV energies, and studies show that neutrinos from this proven 
source can be reconstructed with a high degree of accuracy\citep{Nature}.  Likewise, 
studies have been performed which compare simulated signal events with high-quality 
downgoing muon events \citep{Hodges}.  Because these downgoing events have similar 
properties to the simulated signal events, this provides additional assurance that the 
simulated signal events will have similar properties to the actual signal events we are 
attempting to observe.  Section 3.4 gives a quantitative discussion of systematic errors.
 
To determine the set of data selection criteria that will produce the optimal flux upper 
limit in the absence of a signal, we minimize the Model Rejection Factor (MRF)
\citep{Hill03}.  The MRF is based on the expected detector sensitivity prior to observations: 
\begin{equation}
MRF = \frac{\bar{{\mu}}_{90}({\mathrm{N_{BG,Exp}}})}{{\mathrm{N_{Sig}}}}
\end{equation}
\noindent
where $\bar{{\mu}}_{90}$ is the Feldman-Cousins $90\%$ average event upper limit 
\citep{FeldCous} derived from the expected number of background events 
(N$_{\mathrm{BG,Exp}}$) and N$_{\mathrm{Sig}}$ is the expected number of 
signal events.  N$_{\mathrm{Sig}}$ is determined by convolving the theoretical spectrum 
(${\Phi}$ = dN$_{\mathrm{\nu}}$/dE) with the detector's energy- and angle-dependent 
effective neutrino collecting area ($A_{eff,{\nu}}$) and integrating over the angular 
acceptance of the detector, the energy range of interest ($10^{2}$ to $10^{7}$ GeV), 
and the observation time (assuming 700 bursts contribute equally to the annual expected 
flux):
\begin{equation}
N_{\mathrm{Sig}} = {\int}{\int}{\Phi}(E,{\theta},{\phi}) A_{\mathrm{eff,{\nu}}}(E,{\theta}) {\mathrm{dE d{\Omega} dt}}.
\end{equation}
\noindent
As an intermediate step in the determination of the expected number of signal events, we 
therefore need to determine the detector effective collection area.  
$A_{\mathrm{eff,{\nu}}}$ is determined by the fraction of simulated neutrino events that 
are retained after all data selection criteria are applied.  This area also accounts for 
neutrinos that generate muons passing nearby (but not through) the detector and still cause 
the telescope to trigger.

In determining the optimal data selection criteria for the coincident search, we assume a 
Waxman-Bahcall neutrino spectrum \citep{Waxman}; for the precursor search, we assume a 
Razzaque spectrum \citep{Razzaq}.  In addition to temporal coincidence described 
previously, the most relevant selection criterion for this analysis is the angular mismatch 
(${\Delta}$${\Psi}_{\mathrm{i}}$) between the burst position and the reconstructed event 
track.  This mismatch is determined for each of four separate maximum-likelihood pattern 
recognition algorithms (i = 1 to 4) applied to the timing of the hit OMs (as described in 
Section 2).  The different algorithms are based on different initial seeds and apply a 
different number of iterations to the track reconstruction procedure, thus they are able to 
provide different measures used for discrimination between expected signal and background 
events.  Though they are not completely independent, they do offer improvements to the MRF 
when applied consecutively.  The inherent difference in the muon and neutrino paths, as 
well as the inaccuracies of the reconstruction algorithms, prevent perfect characterization 
of all signal and background events.  Nevertheless, the angular mismatch is quite effective 
as a selection criterion. For example, selecting events with a mismatch angle 
${\Delta}$${\Psi}_{1}$ of less than $12^{\mathrm{o}}$ retains more than $90\%$ of the 
expected signal events, while reducing the background to less than $0.5\%$ (Figure 5).  
Depending upon the changes in the detector characteristics and the analysis tools from year 
to year, the MRF optimization procedure allowed for some variation in the specific track 
reconstruction algorithms applied, as well as the mismatch angle values selected for each 
algorithm (see Table 3).

Several secondary criteria were also used to improve the separation between signal and 
background events.  Included in the secondary criteria is the measured number of hit 
channels---that is, the number of OMs participating in the reconstruction of each event.  
The number of direct hits---hits that occur within -15/+75 ns of the arrival time for light 
propagating from the reconstructed muon track to the OM in question---also serves as a 
useful criterion for data selection.  Direct hits should be due to photons that do 
not scatter, or scatter minimally; their straight trajectories give them a well-defined 
behavior, making them most useful in determining the muon direction.  Additionally, 
the likelihood of a given reconstruction and the angular resolution (${\sigma}_{\Psi}$) of 
the alternate event track reconstruction (the ``paraboloid fit'') provide a useful event 
discriminator, since high quality signal events will have higher likelihoods and superior 
angular resolution compared to the background events.  One additional criterion used in 
this analysis is the uniformity of the spatial distribution of the hit OMs---events with 
hit OMs spread evenly along the track are more likely to be single high-energy 
neutrino-induced muons, whereas events with hit OMs clustered in time and space along the 
track are more likely to be background events.  Different combinations of these criteria 
were applied in the 1997-1999, 2000, and 2001-2003 timeframes, as new analysis tools were 
developed and applied to the GRB neutrino search (see Table 3).

This analysis procedure was applied to bursts with localization errors from the satellite 
observations that are relatively small (typically less than 1$^{\circ}$) and therefore 
inconsequential on the scale of the AMANDA search bin radius.  However, several hundred IPN 
bursts have large localization errors (${\gtrsim}$1/2 of the search bin radius), but still 
lie completely within the field of view of AMANDA.  These were either marginal detections 
near the edge of BATSE's field of view or they were detected by only two IPN 
satellites, which prevents triangulation of their position but allows localization to an 
annular segment.  Eleven of the bursts in the AMANDA dataset are only poorly localized; the 
increased search area for these bursts results in a corresponding increase in the expected 
background rate.  To ensure that this increase does not diminish the overall sensitivity of 
the GRB search, more restrictive selection criteria are applied to these bursts.  Whether 
well localized or poorly localized, each burst has an associated background expected during 
the burst time, calculated from the event rate of the off-time background region multiplied 
by the duration of the time window during which we search for signal events.

The initial criteria were independently selected to optimize the MRF and were then collectively 
optimized in an iterative fashion.  The optimal criteria depended on the zenith angle of 
the burst, due to the higher observed background rate for bursts closer to the horizon.  
The criteria for higher background rates (i.e. low zenith angle bursts) were also applied 
to bursts with large satellite localization errors, regardless of the actual zenith angle 
of the burst.  Table 3 lists all data selection criteria used for the year-by-year GRB 
analyses, as well as the selection criteria for the precursor search applied in 2001-2003.  
Though the data selection criteria are optimized for specific models of neutrino emission,  
other models can also be tested using the Green's Function Fluence Limit Method (see 
Results).  While the muon track reconstruction algorithm is very accurate, there is a 
small probability that a downgoing muon will be misreconstructed in the upgoing direction; 
such events are the primary background for the GRB search.  After the application of data 
selection criteria, background events have an observed rate of ${\sim}5 {\times} 10^{-5} 
{\mathrm{Hz}}$ (with some seasonal variation).  
Figure 6 show the effective area for neutrinos for the AMANDA-II detector after all data 
selection criteria are applied.
Due to the large instrumented area and modest background rejection requirements of this 
analysis, AMANDA-II has an $A_{eff}$ significantly larger than any other 
contemporaneously-operating neutrino detector (e.g. Baikal \citep{Baikal}, SuperKamiokande 
\citep{SuperK}, and SNO \citep{SNO}).  A determination of the relative MRF for a 
subset of bursts from the year 2000 analysis is shown in Figure 7 (the arrow indicates the 
MRF for the selected criteria).

Prior to ``unblinding'' the analysis and determining the number of events we observe, we 
determine the flux sensitivity to simulated GRB neutrinos.  Results from the 268 bursts 
observed from 1997 to 1999 have been presented previously \citep{Bay, Rellen}.  We combine 
these initial observations with the results from the analysis of 151 bursts in the data 
collected in 2000-2003.  The flux sensitivity for all 419 bursts is the MRF prior to 
observations (see equation 5) multiplied by the normalization of the input spectrum; that 
is, 
$E^{2}$${\Phi}_{\mathrm{\nu}}$ ${\leq}$2 ${\times}$ $10^{-8}$ ${\mathrm{GeV cm^{-2} s^{-1} 
sr^{-1}}}$ for a Waxman-Bahcall muon neutrino spectrum with 90\% of the events expected 
between $\sim$10 TeV and $\sim$3 PeV.  This sensitivity is calculated prior to the inclusion of 
systematic uncertainties.

\subsection{Uncertainties in Observation and Modeling}

There are several potential sources of systematic uncertainty in this analysis, including 
the Monte Carlo simulations of signal events, the modeling of the scattering and absorption 
lengths of the South Pole ice, and the OM response to incident photons.  For the flux upper 
limits incorporating IPN bursts, the potential for inclusion of bursts which do not fit 
models based upon BATSE triggered bursts contributes to the overall uncertainty as well.  
Additionally, some bursts are of unknown duration--for the purposes of this search, they 
were classified as long-duration bursts so that we would not needlessly exclude any 
possible signal events.  However, including all such bursts will potentially overestimate 
the signal event predictions for models based solely upon long-duration bursts.  Finally, 
previous results from 1997-1999 were applied only to the Waxman-Bahcall model; limitations 
in the simulation procedures in place at that time means that adapting these results to 
other models will introduce uncertainties in the expected neutrino event rate.

The scattering and absorption lengths of the ice were measured during the 1999-2000 austral 
summer with in situ lasers and LED flashers \citep{Kurt}.  While these measurements were 
extremely accurate, the limited precision with which they were implemented in our detector 
simulations contributes about $15\%$ to the overall uncertainty.  Furthermore, the quantum 
efficiency of the photomultiplier tubes is known to within $10\%$, while the transmission 
efficiency of the glass pressure housing and the optical gel is known to a comparable 
precision.  However, triggering depends on the detection of photons by 24 or more PMTs, so 
the uncertainty in a single OM does not translate directly into an uncertainty in the 
expected flux.  Detailed simulations show that the quantum and transmission efficiencies 
together contribute only about 7\% uncertainty in the expected neutrino flux \citep{Reco}.  
Though the GRB search implements a different methodology from other IceCube analyses 
\citep[e.g. the point source search detailed in][]{Elisa}, the values for the individual 
contributions to the uncertainty are consistent across these different analyses.

Additionally, a statistical correction is required when IPN bursts are incorporated into 
the flux upper limits for models initially based on BATSE observations.  In principle, 
BATSE has a sensitivity comparable to the suite of other IPN satellites treated 
collectively; observationally, their duration distribution seems qualitatively to be 
derived from the same bimodal population (Figure 8).  However, the characteristics of the 
bursts detected by satellites with different sensitivities are not completely identical.  
BATSE non-triggered bursts have on average less than 1/10 of the peak photon flux of their triggered 
counterparts, and if we assume that the neutrino flux scales as the photon flux, then 
including non-triggered bursts in the upper limit calculation would artificially 
increase the expected number of signal events, and thus lead to a flux upper limit that is 
too restrictive.  We calculate (see Appendix A) that 12\% of the IPN bursts should not be 
considered equivalent to BATSE triggered bursts, and thus should be excluded from the 
dataset.  This leads to a 3\% correction in the number of expected signal events.  
Furthermore, for models based solely on long-duration bursts such as \citep{Murase, 
Razzaq}, the inclusion of bursts of unknown duration may also lead to an overestimation of 
the number of expected signal events.  In Appendix A, we derive a statistical correction of 
6\% to the expected number of signal events due to this effect.

Finally, we determine the uncertainty introduced when the previous results from 1997-1999 
are applied to theoretical predictions other than the Waxman-Bahcall model.  Though the 
uncertainties specifically for the Waxman-Bahcall model are well understood and are 
incorporated into the previous results, limitations in the simulation procedures at the 
time of the previous analysis lead to a further uncertainty in the neutrino event rate for 
the Murase-Nagataki and Razzaque et al. models of $\sim$20\%.  When we combine the results 
from the 268 bursts from 1997-1999 with the results from 151 bursts from 2000-2003 into a 
single flux upper limit, we assume conservatively that the neutrino event rate for the 
bursts from 1997-1999 is overestimated by 20\%.  

All significant sources of uncertainty for the GRB analysis, along with the correction 
factors, are summarized in Table 4.  While the reduction in the expected neutrino event 
rate for the 1997-199 bursts is not specifically enumerated in this Table, it is 
incorporated into the relevant flux upper limits discussed in the next section.  Assuming 
no correlation among the other uncertainties, we summed the different factors in quadrature 
and applied the other relevant corrections to obtain a total uncertainty of +16\%/-17\% 
(+15\%/-18\% for models based on long-duration bursts only) in the total detector exposure, 
and therefore in the number of signal events and the flux and fluence upper limits.  This 
is comparable to the uncertainty determined by \citep{Hodges}, who also characterized the 
agreement between the simulated signal events and high-quality downgoing muon events, which 
served as a proxy for the expected signal events for AMANDA analyses.

\section{Results} \label{bozomath}

We observe zero events from the 419 Northern Hemisphere bursts searched during the years 
1997 to 2003, which is consistent with the background estimate of 1.74 events (Table 
5)\footnote{We also searched for neutrino emission from 153 additional non-triggered bursts 
discovered in the BATSE archival data; we observed zero events from these bursts as well.  
We do not include these results in the flux upper limits or MRF determinations.}.  Since 
the observed number of events is less than the expected background, the flux upper limits 
for the coincident muon neutrino search is approximately a factor of three better than the 
expected sensitivity (i.e. the observed MRF for a Waxman-Bahcall flux is 1.3 compared to 
the expected value of 3.8).  Figure 9 shows the 90\% C.L. flux upper limits relative to the 
Waxman-Bahcall, Razzaque, and Murase-Nagataki models.  Though our analysis was restricted 
to bursts located in the Northern Hemisphere (2${\pi}$ sr), all flux upper limits are for 
the entire sky (4${\pi}$ sr).  Including the systematic uncertainties in the manner 
outlined by \citet{Conrad}, we calculate the coincident muon neutrino flux upper limit for 
the Waxman-Bahcall spectrum to have a normalization at 1 PeV of 

\noindent
$E^{2}$$\Phi_{\mathrm{\nu}}$ ${\leq}$ 6.0 $\times$ $10^{-9}$ $\mathrm{GeV cm^{-2} s^{-1} sr^{-1}}$,

\noindent
with 90\% of the events expected between $\sim$10 TeV and $\sim$3 PeV.

We place similar constraints on the model parameters of \citet{Murase}.  Based on our null 
result, Parameter Set C is highly disfavored for all variations in their parameters, though 
this particular set is disfavored on other grounds as well, and is only briefly described 
in their work.  Parameter Set A is ruled out (MRF=0.82) by the current AMANDA 
observations at the 90\% confidence level.  However, it is important to note that Parameter 
Set A uses a baryon loading factor that is fine-tuned to provide significant neutrino flux.  
Other, possibly more realistic, values for the baryon loading would significantly reduce 
the expected neutrino emission, and therefore result in an MRF that is higher by an order 
of magnitude or more.  The original model incorporates only long-duration bursts that 
follow the cosmic star-formation rate \citep{Nagataki}; incorporating all short bursts 
would yield flux upper limits that are better than those presented here by approximately 
13\%, which includes removing the ``correction'' due to incorporating of bursts of unknown 
duration (see Section 3.4).

Our combined results from 1997-2003 also constrain the supranova model of Razzaque et al.  
We begin by considering the assumption that all GRBs are preceded by supernovae that 
produce a circumburst environment ideally suited for neutrino production.  The observed MRF 
for this case is 0.40, and thus we exclude the predicted neutrino flux at the 90\% level.  
Furthermore, the flux upper limit determined for this model is derived from observations of 
long bursts only.  As with the results of \citet{Murase}, if this model is expanded to 
include short-duration bursts, the flux upper limit improves by approximately 13\%.  
However, only a very small number of all bursts (${\sim}$4 out of many thousands) have been 
observed in association with SNe.  And, as described in Section 2, at least a fraction of 
these SNe did not occur at an ideal time relative to the burst.  Thus, AMANDA's results 
confirm previous observations that lead us to expect less than maximal emission from this 
model of GRB neutrino production.

Finally, we observe zero events (on an expected background of 0.2 events) from the 
precursor time period of the bursts from 2001-2003 (Table 6).  The precursor model of 
neutrino production was tested for only a small subset of the long-duration bursts, and the 
neutrino energy spectrum peaks at a level where the AMANDA-II sensitivity is greatly 
reduced.  Thus, the flux upper limit for the precursor model is significantly less 
restrictive.

The results of these analyses can also be applied to any other hypothesized spectrum by 
using the Green's Function Fluence Limit formula, in a method similar to that presented by 
the Super-Kamiokande Collaboration \citep{SuperK}.  By folding the energy-dependent 
sensitivity of the detector into a desired theoretical spectrum, one can straightforwardly 
calculate a flux upper limit for that specific spectrum.  The Green's Function fluence 
upper limit for AMANDA-II (Figure 10) extends several orders of magnitude in energy beyond 
the range of the Super-Kamiokande limit, and is approximately an order of magnitude lower 
than the Super-Kamiokande results in the region of overlap, primarily due to the much 
larger effective area of AMANDA-II.  For example, at 100 TeV we calculate 
$F_{\mathrm{\nu}}$ ${\leq}$ 1.7 $\times$ $10^{-7}$ $\mathrm{cm^{-2}}$ (see also Appendix 
B).  As this method does not rely on averaging burst properties (as many specific models 
do), it is particularly effective for incorporating large burst-to-burst variations in 
expected muon neutrino flux \citep[e.g. for GRB030329, see][]{MikeSwift}.

\section{Conclusion and Outlook}

The AMANDA dataset has been searched for muon neutrino emission from more than 400 GRBs 
based on temporal and spatial coincidence with photon detections from numerous other 
observatories.  We determined that the detector was operating in a stable fashion during 
all of these bursts, and we have shown that the application of a number of data selection 
criteria lead to an optimized value of the Model Rejection Factor for the Waxman-Bahcall 
neutrino spectrum.  After the application of these criteria, zero neutrino events were 
observed in coincidence with the bursts, resulting in the most stringent upper limit on the 
muon neutrino flux from GRBs to date.  We have compared this limit to the flux predictions 
from several prominent GRB models based on averaged burst properties.  We constrain the 
parameter space of a number of these models at the 90\% confidence level; in particular, 
our flux upper limit is more than a factor of 2 below the most optimistic predictions of 
Razzaque et al.  However, we do not yet rule out the predictions of the canonical Waxman \& 
Bahcall model.  Furthermore, because individual bursts vary significantly in their expected 
neutrino spectra, we have presented a spectrum-independent method for determining flux 
upper limits for these bursts.  The observations detailed in this work will play a 
significant role as future analyses seek to further constrain various theoretical models.

Finally, AMANDA's search for muon neutrinos from more recent GRBs will benefit greatly from 
the advanced capabilities of the Swift satellite \citep{Swift}, as will the GRB searches 
of other neutrino observatories currently in operation \citep{Baikal,Nestor,Antares}.  
While Swift's rate of GRB detections is lower than that of BATSE, the spatial localizations 
of the bursts by Swift are much more precise, which will obviate the need for a special 
analysis of poorly-localized bursts with its accompanying reduction in signal detection 
efficiency.  Additionally, the InterPlanetary Network of satellites will continue to operate, 
and will incorporate newer instruments as they come online.  In particular, future missions 
such as the Gamma-Ray Large Area Space Telescope \citep{GLAST} will provide an even greater 
number of GRB localizations for use in neutrino searches.  Furthermore, while analyses similar 
to the one presented here will continue to search specifically for muon neutrino flux in 
coincidence with photon observations of gamma-ray bursts, the method described here can be 
expanded to search for neutrinos correlated with other transient point sources as well (see 
Appendix C).  In the future, AMANDA and its successor, IceCube, will have many more 
opportunities to detect neutrino emission from a host of astrophysical sources.  Construction 
of IceCube is currently underway, and the instrumented volume for the partial detector is already 
significantly larger than the final instrumented volume of AMANDA.  A fully-instrumented 
IceCube detector should surpass AMANDA's flux upper limits within its first few years of 
operation.

\acknowledgments

\section{Acknowledgements}

The authors are thankful to E. Waxman and K. Murase for productive and thought-provoking discussions, S. Desai for detailed discussions regarding the 
Super-Kamiokande GRB analysis, and the additional members of the Interplanetary Network for providing data that were crucial to this work.  The IceCube 
collaboration acknowledges the support from the following agencies: National Science Foundation-Office of Polar Program, National Science 
Foundation-Physics Division, University of Wisconsin Alumni Research Foundation, Department of Energy, and National Energy Research Scientific 
Computing Center (supported by the Office of Energy Research of the Department of Energy), the NSF-supported TeraGrid system at the San Diego 
Supercomputer Center (SDSC), and the National Center for Supercomputing Applications (NCSA); Swedish Research Council, Swedish Polar Research 
Secretariat, and Knut and Alice Wallenberg Foundation, Sweden; German Ministry for Education and Research, Deutsche Forschungsgemeinschaft (DFG), 
Germany; Fund for Scientific Research (FNRS-FWO), Flanders Institute to encourage scientific and technological research in industry (IWT), Belgian 
Federal Office for Scientific, Technical and Cultural affairs (OSTC); the Netherlands Organisation for Scientific Research (NWO); M. Ribordy 
acknowledges the support of the SNF (Switzerland); J. D. Zornoza acknowledges the Marie Curie OIF Program (contract 007921).
K. Hurley is grateful for IPN support under the following contractsand grants: Ulysses, JPL958056; Mars Odyssey, JPL 1282043; HETE, 
MIT-SC-R-293291; the U.S. BeppoSAX Guest Investigator program; the U.S. INTEGRAL Guest 
Investigator program.  Integration of NEAR into the Interplanetary Network was supported by 
NASA's NEAR Participating Scientist program.  Integration of RHESSI was supported by NASA's 
Long Term Space Astrophysics program under grant NAG5-13080.

{\it Facilities:} \facility{AMANDA}

\appendix

\section{Model-Dependent Statistical Corrections to Flux Upper Limits}

Though the ${\nu}$ flux formulation of \citet{Waxman} explicitly links GRB neutrinos to the 
UHECR flux, elsewhere a formulation based on BATSE observations is treated in a comparable 
fashion, and is considered to arise from the same underlying phenomena \citep{WaxBah}.  
Thus it is necessary to address the limitations introduced by AMANDA's reliance upon BATSE 
observations.  As described in Section 3.4, models defined initially in terms of BATSE 
observations were also applied to bursts detected by the other IPN satellites.  However, we 
cannot assume that characteristics of bursts detected by satellites with different 
sensitivities are completely identical.  Since BATSE was decommissioned in May of 2000, 
there is no longer a way to cross-correlate the two datasets.  Non-triggered BATSE bursts 
have on average less than 1/10 of the peak photon flux of the triggered bursts; assuming 
that the energy of neutrinos scales with the energy carried by gamma rays, we expect only a small 
fraction of the standard neutrino flux from these non-triggered bursts.  Thus, if 
non-triggered bursts are inadvertently included in the flux upper limit, they will 
artificially improve that limit, because the extra bursts are assumed to have a larger 
neutrino flux than they would actually possess.   

During the period of simultaneous operation from 1991 to 2000, 1088 IPN bursts were 
observed by BATSE, 953 of which were triggered.  Undoubtedly some of these bursts did not 
trigger BATSE for reasons other than a lower flux.  For example, BATSE may have been 
powered down, may have been in the vicinity of the South Atlantic Anomaly, or may have 
experienced unrelated on-board performance problems.  However, we assume conservatively 
that all such bursts did in fact exhibit the lower flux common to non-triggered bursts.  
Therefore, ${\sim}$12$\%$ of the IPN bursts should not actually be a part of the dataset 
that is compared with the models that are based upon BATSE's triggered GRB rate.  Because 
IPN bursts are expected to contribute ${\sim}$25\% of our detectable signal, this effect 
reduces the total expected neutrino flux by ${\sim}$3\%.  This correction is applied 
asymmetrically to the overall uncertainty, because it can hinder, but not improve, the 
effectiveness of the analysis (see Table 4).

For models based solely on long-duration bursts, such as \citep{Murase, Razzaq}, the inclusion 
of bursts of unknown duration may also lead to an overestimation of the expected signal 
events, and thus a flux upper limit that is too restrictive.  In order to ensure that we 
would not exclude potentially detectable neutrino events, the 75 bursts of unknown duration 
included in the dataset are assumed to last 100 s (for 1997-1999) or 50 s (for 2000-2003).  
Thus, for purposes of data analysis, they are classified as long-duration bursts.  However, 
this necessitates a statistical correction to the resulting flux limits.  We assume that up 
to 1/3 of these bursts may in fact be short-duration, based upon the standard ratio of 
short- to long-duration bursts observed by BATSE.  So, of the 389 bursts known (or assumed) 
to be long-duration, 25 were excluded from the relevant limits, thus reducing the expected 
number of signal events by 25/389, or ${\sim}$6\%.  This correction is likewise applied 
asymmetrically to the overall uncertainty.

\section{Green's Function Fluence Upper Limit Calculation}

We show here sample calculations of the differential neutrino fluence upper limit, as well 
as a procedure to determine the integrated fluence and flux upper limits, following the 
Green's Function method set out in Section 3 of \citet{SuperK}.  The fluence upper limit 
calculation assumes a monochromatic neutrino spectrum; the calculation is repeated at 
different values of the neutrino energy.  The benefit of this method is that an integrated 
fluence upper limit can then be determined for any input spectrum, whether it be based on 
all of the bursts in this dataset or only on a subset of all bursts.

The fluence upper limit is defined as
\begin{equation}
F(E) \leq \frac{{\mathrm{N_{90}}}}{{\mathrm{A_{eff,{\nu}}}}({\mathrm{E_{\nu}}})}
\end{equation}
\noindent
where ${\mathrm{N_{90}}}$ is ${\mathrm{{\mu}_{90}}}$/${\mathrm{N_{Bursts}}}$ and 
${\mathrm{A_{eff,{\nu}}}}$ is the energy-dependent neutrino effective collecting area (see 
Section 3.3)\footnote{Instead of using the neutrino effective area, one could also use the 
muon effective area multiplied by the neutrino to muon conversion probability \citep[as 
in][]{SuperK}; in the case of AMANDA one must also account for attenuation of neutrinos in 
the earth.}.

Figure 10 is determined by the results of AMANDA's 2000-2003 observations.  For example, 
${\mathrm{{\mu}_{90}}}$ = 1.30 and ${\mathrm{N_{Bursts}}}$ = 151; therefore, 
${\mathrm{N_{90}}}$ = 8.61 $\times$ 10$^{-3}$.

For $E_{\mathrm{{\nu}}}$ = 100 TeV (near the peak of the predicted neutrino flux), 
$A_{\mathrm{{eff,{\nu}}}}$ = 5.0 $\times$ 10$^{4}$ cm$^2$ therefore F(100 TeV) ${\leq}$ 1.7 
$\times$ 10$^{-7}$ cm$^{-2}$.

We now determine the integrated fluence upper limit explicitly for an E$^{-2}$ spectrum, as 
well the Waxman-Bahcall spectrum.  First, the integrated fluence, ${\mathrm{F_{int}}}$ for 
an E$^{-2}$ spectrum is 
\begin{equation}
F_{\mathrm{int}} \leq [\int_{250 {\mathrm{GeV}}}^{10^{7} {\mathrm{GeV}}} \frac{C 
E_{\mathrm{\nu}}^{-2}}{F(E)} {\mathrm{dE_{\nu}}}]^{-1} =
1.4 {\times} 10^{-5} {\mathrm{cm^{-2}}},
\end{equation}
\noindent
where C is the factor required to normalize the neutrino spectrum to unity---in this case, 
C = 250 GeV.  This integrated fluence upper limit is significantly lower than the results 
of similar calculations performed by \citet{SuperK} (we combine the ${\mathrm{{\nu}_{\mu}}}$ 
and ${\mathrm{\bar{{\nu}}_{\mu}}}$ fluences into a single limit, while they present two 
separate fluence upper limits).  However, a direct, quantitative comparison between these 
two results cannot be made due to the vastly different energy ranges of the two instruments.  
Note also the limits of integration employed here---though AMANDA is sensitive to neutrinos 
at higher and lower energies, the vast majority of the flux from GRBs is expected to come 
from neutrinos of a few hundred GeV to a few PeV.

Now we determine the integrated fluence upper limit for the Waxman-Bahcall spectrum, to 
provide a further example of the wide applicability of the Green's Function method:
\begin{equation}
F_{\mathrm{int}} \leq [\int_{250 {\mathrm{GeV}}}^{10^{5} {\mathrm{GeV}}} 
\frac{C E_{\mathrm{\nu}}^{-1} E_{\mathrm{Break}}^{-1}} {F(E)} {\mathrm{dE_{\nu}}}
+ \int_{10^{5}}^{10^{7}} \frac{C E_{\mathrm{\nu}}^{-2}}{F(E)} {\mathrm{dE_{\nu}}}]^{-1}
= 5.3 {\times} 10^{-7} {\mathrm{cm^{-2}}},
\end{equation}
\noindent
where C again is the constant required to normalize the overall spectrum to 
unity; here C = $7.0 {\times} 10^{-5} {\mathrm{GeV}}$.

Finally, we compare this fluence upper limit to the flux upper limit derived for the 
Waxman-Bahcall spectrum in Section 4.  To do this, we must convert the integrated 
fluence upper limit into a differential all-sky flux upper limit per burst; that is, from 
units of cm$^{-2}$ to units of $\mathrm{GeV^{-1} cm^{-2} s^{-1} sr^{-1}}$:

\begin{equation}
\frac{F_{\mathrm{int}}}{{\Omega} {\mathrm{t}}} = \frac{5.3 {\times} 10^{-7}} {(4{\pi}) 
(3.15 {\times} 10^{7}/700)}= 9.4 \times 10^{-13} 
{\mathrm{cm^{-2} s^{-1} sr^{-1}}}.
\end{equation}

Next, we multiply by the normalization of the energy spectrum and take the differential to 
provide a flux upper limit of

\noindent
E$^{2}$${\Phi}_{\mathrm{\nu}}$ ${\leq}$ 1.3 ${\times}$ 10$^{-8}$ $\mathrm{GeV cm^{-2} 
s^{-1} sr^{-1}}$.

\noindent
This is nearly identical to the flux upper limit derived in the manner described in Section 
3 for 151 bursts from 2000 to 2003 (see also Table 5, where an MRF of 2.5 yields a flux 
upper limit of E$^{2}$${\Phi}_{\mathrm{\nu}}$ ${\leq}$ 1.1 ${\times}$ 10$^{-8}$ 
$\mathrm{GeV cm^{-2} s^{-1} sr^{-1}}$, consistent with the result derived above to within 
the applicable uncertainties).

Thus we show that the Green's Function Method agrees with calculations which explicitly 
incorporated prior assumptions about the GRB neutrino spectrum.  Therefore, this alternate 
method provides a powerful tool for determining the flux upper limit based on AMANDA 
observations for any proposed neutrino spectrum.

\section{Expanding the GRB Search to Other Transient Point Sources}

While this work has provided the most stringent upper limit to date specifically for muon 
neutrino flux for gamma-ray bursts in coincidence with photon observations, the method 
described above can be expanded to search for other transient point sources as well.  X-ray 
flares occurring minutes to hours after a GRB are thought to be caused by re-activation of 
the GRB central engine, and are a natural candidate for correlated neutrino searches 
\citep{Murase2}. Additionally, photon emission from supernovae could 
be used as a key element in searches for neutrino emission from jet-driven supernovae and 
${\gamma}$-ray dark (``choked'') GRBs \citep{Razza2}.  Jet-driven supernovae are expected 
to accelerate baryonic material to mildly relativistic energies (the Lorentz boost 
${\Gamma}$ ${\sim}$ a few), which may subsequently result in significant neutrino emission 
\citep{Ando}.  Not all supernovae will be jet-driven, but population estimates vary between 
0.2\% and 25\% of all type Ib/c SNe \citep{Putten,Berger,Soder}.  Given the number of such 
supernovae observed annually, it is reasonable to search for a neutrino signal from these 
events.

Another reason to search for neutrino emission from supernovae becomes apparent when we 
consider the recently-established SN-GRB connection.  Several supernovae (including 1998bw 
and 2003dh) are known to be associated with GRBs.  Furthermore, Razzaque et al. (2003b) 
describe a scenario where as many as $10^{3}$ times the standard number of GRBs occur, 
though in these bursts the photon jet does not succeed in escaping the stellar envelope 
(the ${\gamma}$-ray dark GRBs).  For these types of bursts, no gamma-rays will be observed.  
However, if even a fraction of these GRBs are associated with SNe (the fraction for observed 
GRBs has been calculated to be in the range of $10^{-2}$ to $10^{-3}$ \citep{Bissaldi}), 
then it will be possible to search for neutrinos in the time period surrounding the SN 
emission (provided the SN start time, the GRB time delay relative to the SN, and the 
duration of the GRB can be estimated with sufficient precision).  Because these SNe are 
localized transient phenomena, the primary selection criteria for the GRB analysis (spatial 
and temporal correlation) are an excellent starting point for such a search, though it is 
possible that not all of the other data quality cuts used in the GRB search would be 
optimal for a supernova search.  Finally, it is also possible to complement any of the 
transient point source searches described above by inverting the search algorithm, that is, 
by implementing Target of Opportunity photon searches based on spatio-temporal localization 
of potential neutrino events \citep{Marek}.  Any of these searches can potentially be of 
great benefit to the long-term goals of multi-messenger astronomy.

\clearpage

\begin{figure}
\epsscale{1.0}
\plotone{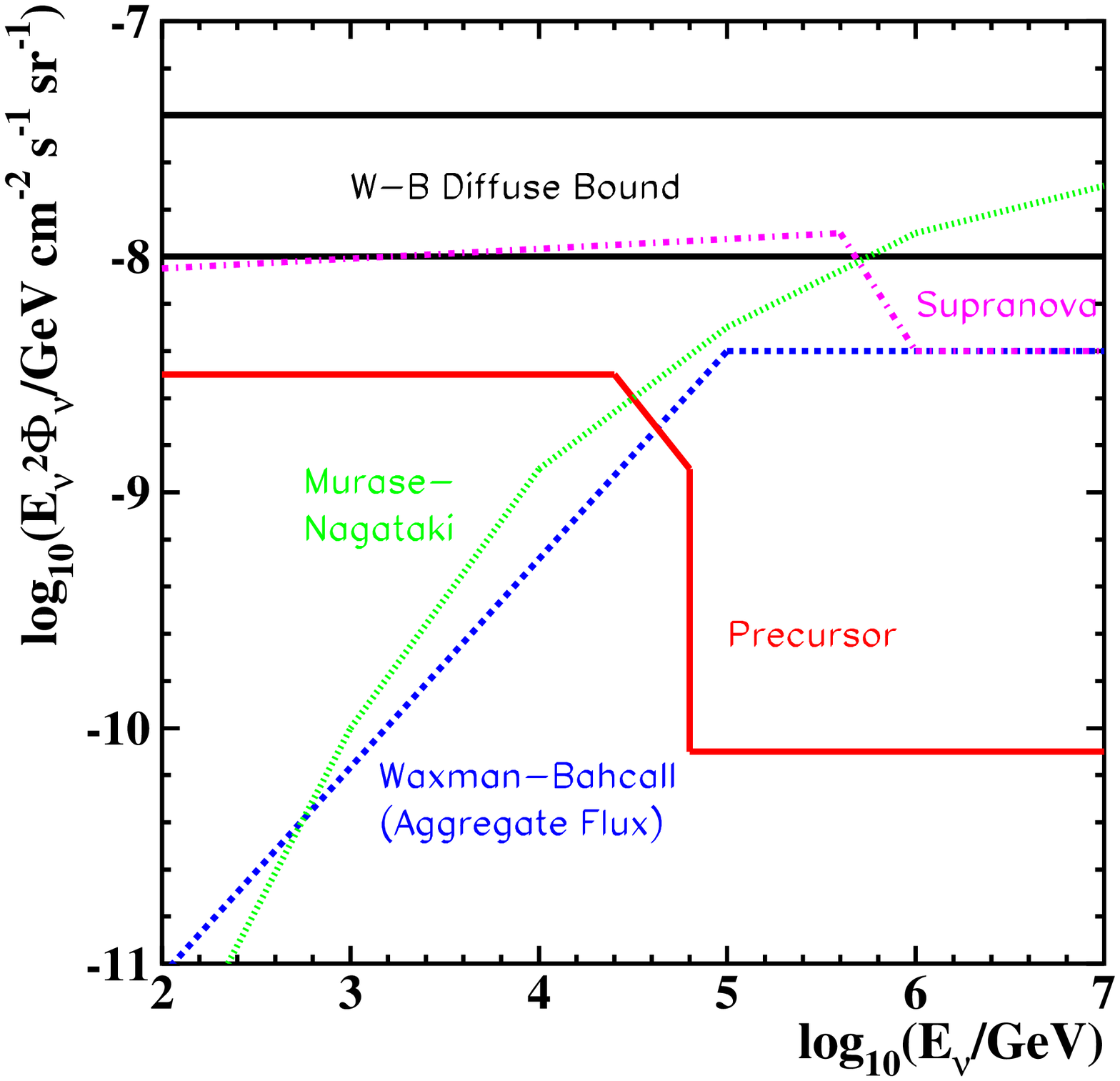}
\caption{Predicted differential muon neutrino flux as a function of energy for four 
different models of GRB neutrino production: the precursor model (solid line), the 
canonical Waxman-Bahcall model (thick dotted line), the Murase-Nagataki model (thin dotted 
line), and the supranova model (dot-dashed line).  All models include the effect of 
${\nu}$ oscillations.  The diffuse neutrino bounds determined from cosmic ray observations 
with (upper horizontal line) and without (lower horizontal line) z evolution are also shown 
for reference.
\label{fig1}}
\end{figure}

\begin{figure}
\epsscale{1.0}
\plotone{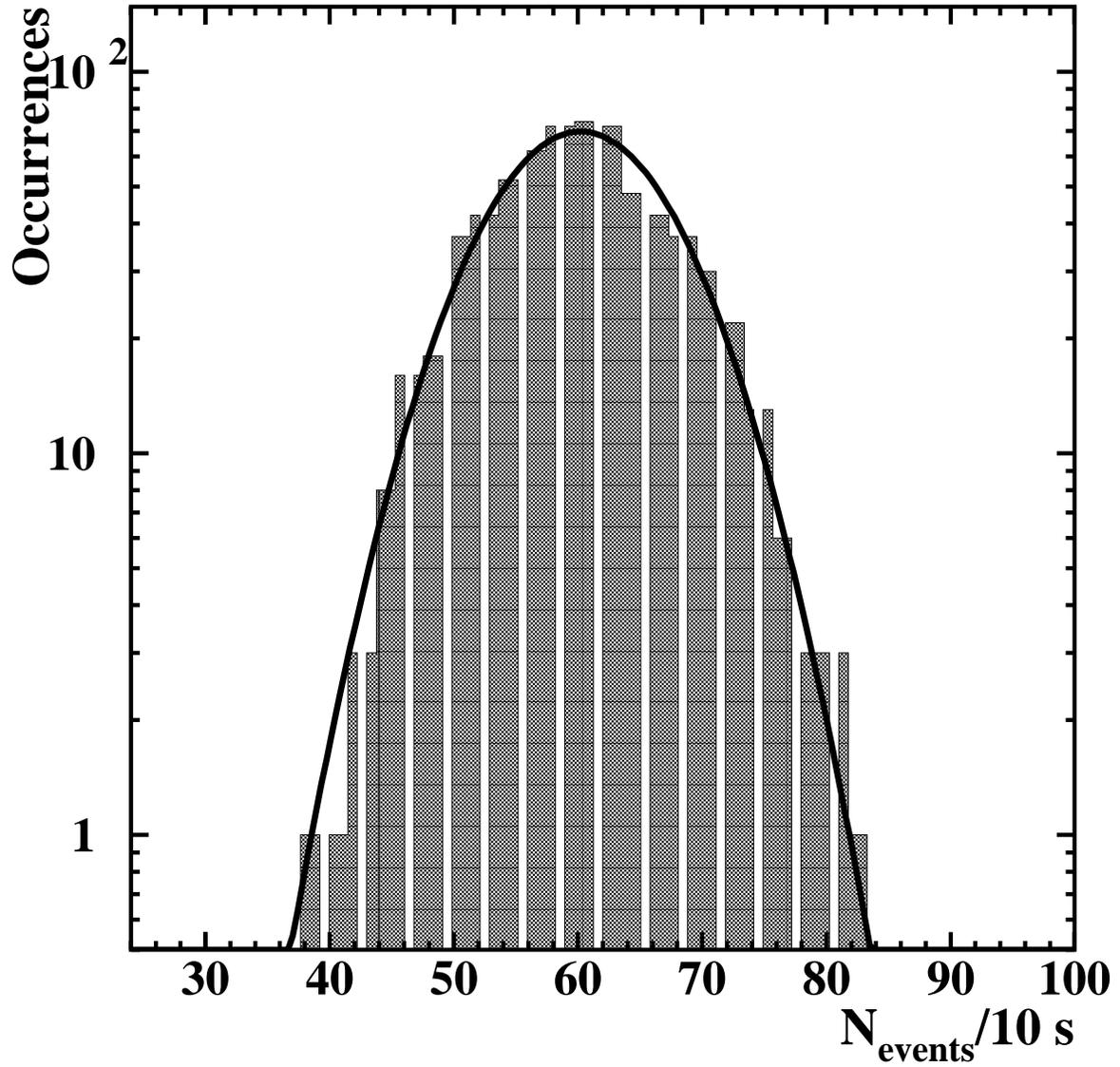}
\caption{A stable period of detector activity, shown by the nearly Gaussian random temporal 
distribution of events in each 10-second bin during the off-time period of BATSE GRB 
6610.  Initial selection criteria have been applied to these data, but the GRB-specific criteria 
have not yet been applied.
\label{fig2}}
\end{figure}

\begin{figure}
\epsscale{1.0}
\plotone{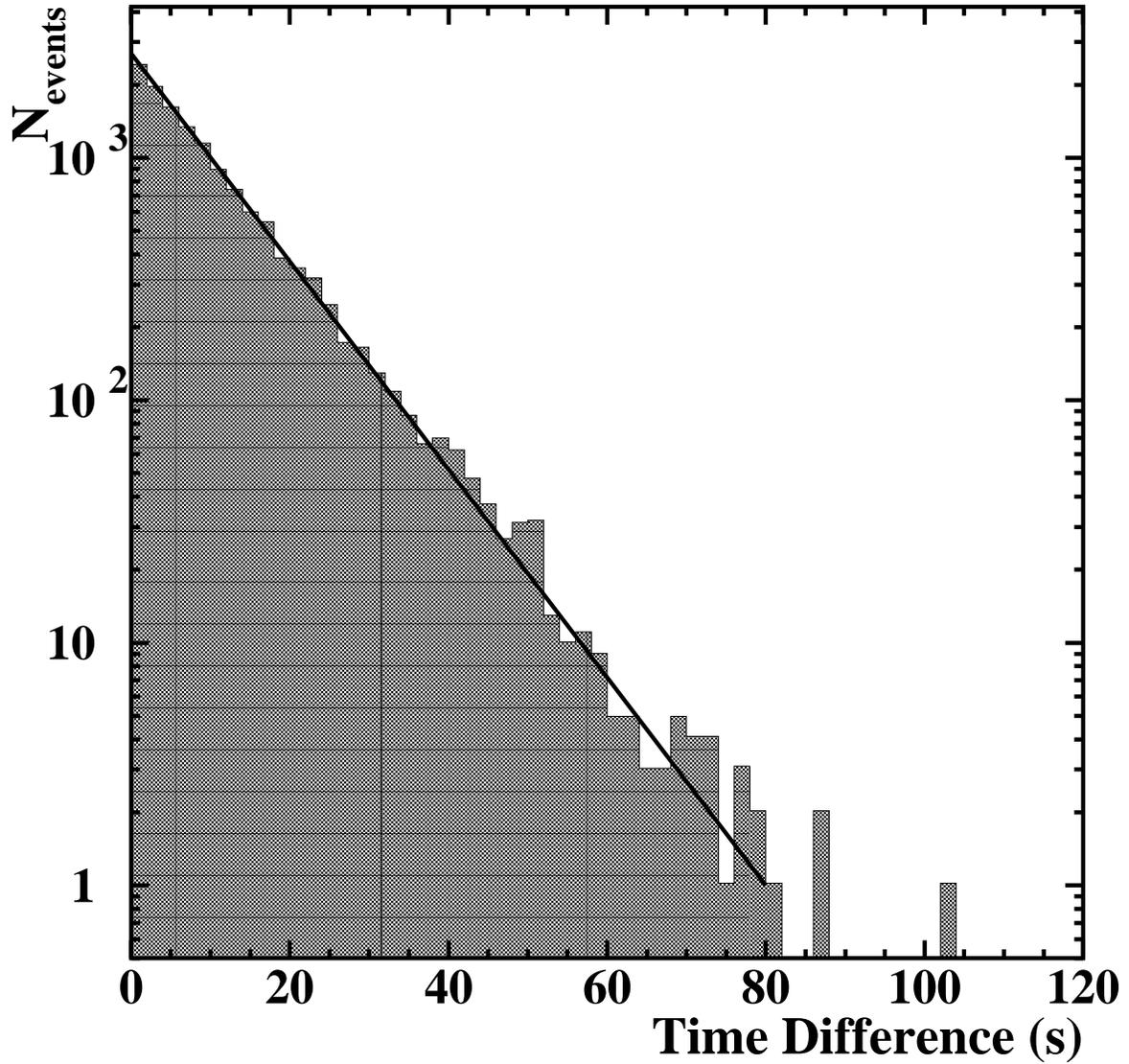}
\caption{Time difference ($\mathrm{{\delta}t}$) between subsequent events during the 
background time period of a representative GRB, after application of initial data quality 
cuts.  There is no evidence for significant gaps in the data that could produce a ``false 
negative'' result.
\label{fig3}}
\end{figure}

\begin{figure}
\epsscale{1.0}
\plotone{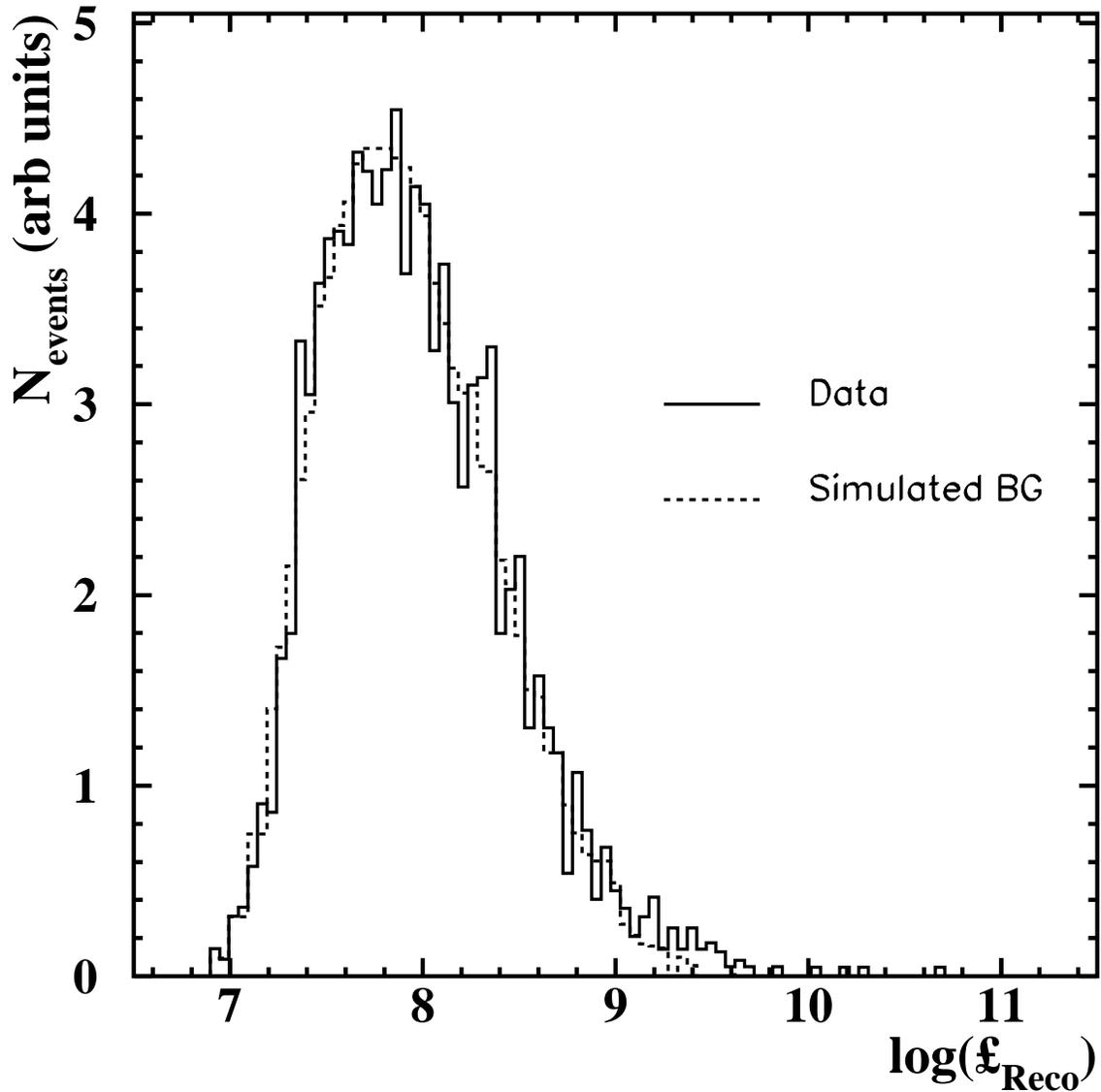}
\caption{A comparison of the likelihood of track reconstruction, 
${\mathrm{{\pounds_{reco}}}}$ for observed data (solid line) and simulated background 
events (dashed line).  Both curves are normalized after preliminary data selection criteria 
are applied.  The close agreement signifies that our simulations are properly modeling the 
observed events, thus providing additional evidence for the trustworthiness of the 
simulated signal events as well.
\label{fig4}}
\end{figure}

\begin{figure}
\epsscale{1.0}
\plotone{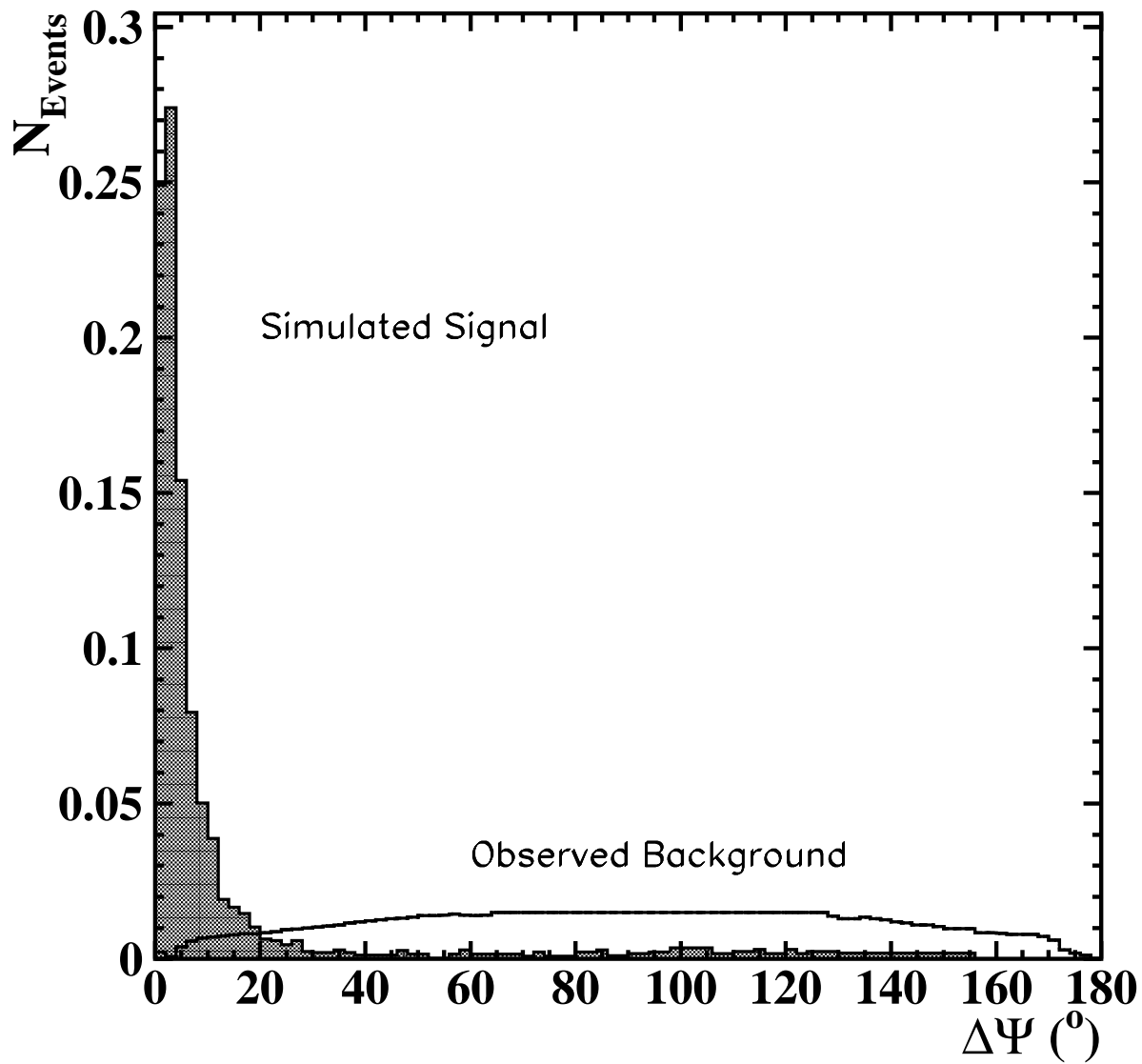}
\caption{The expected distribution of angular mismatch ${\Delta}$${\Psi}_{1}$ for a simulated 
muon neutrino spectrum (shaded region) and observed background (open region).  
${\Delta}$${\Psi}_{1}$ = 0 is the position of the burst determined from photon 
observations.  Selecting events with ${\Delta}$${\Psi}_{1}$${\leq}$12$^\circ$ 
retains more than $90\%$ of the signal events.
\label{fig5}}
\end{figure}

\begin{figure}
\epsscale{1.0}
\plotone{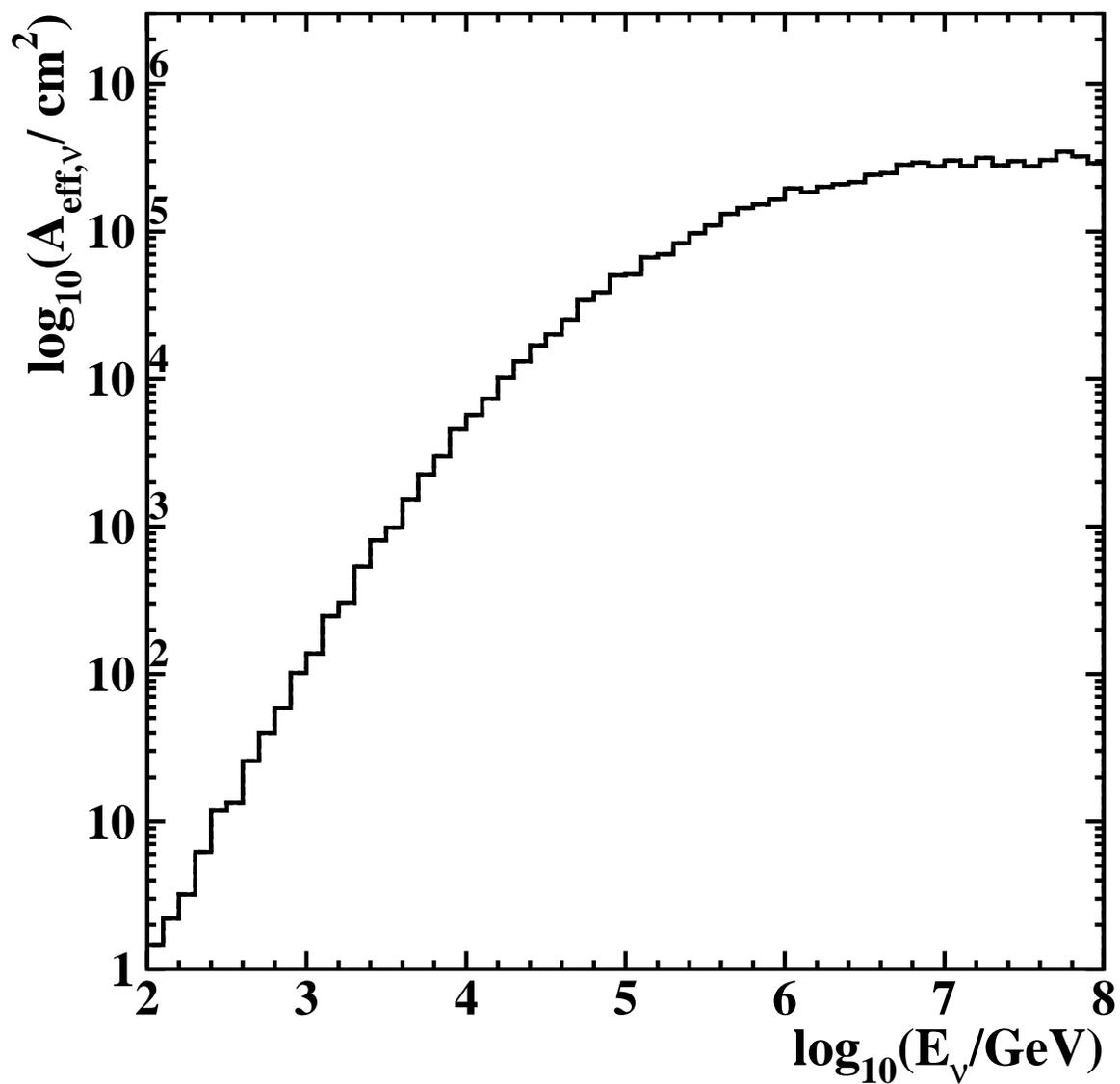}
\caption{Angle-averaged muon neutrino effective area for the AMANDA-II (years 2001-2003) 
coincident search algorithm, based upon Monte Carlo simulations of expected signal 
events from the Northern hemisphere.
\label{fig6}}
\end{figure}

\begin{figure}
\epsscale{1.0}
\plotone{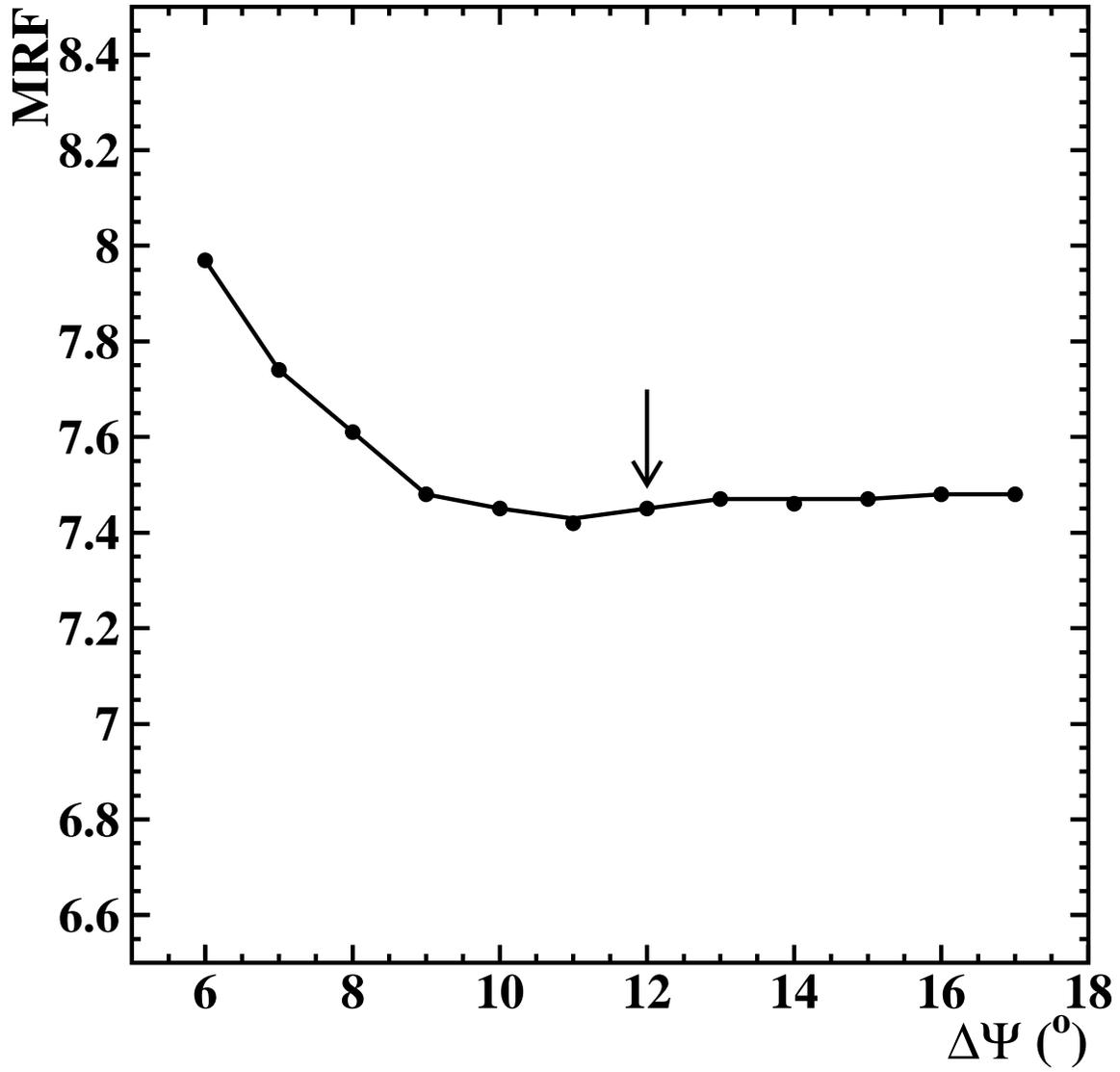}
\caption{Relative Model Rejection Factor (MRF) as a function of angular mismatch 
(${\Delta}$${\Psi}_{1}$) between the burst position and the reconstructed track, for the subset 
of bursts from 2000.  The arrow indicates the mismatch angle selected for this analysis.
\label{fig7}}
\end{figure}

\begin{figure}
\epsscale{1.00}
\plotone{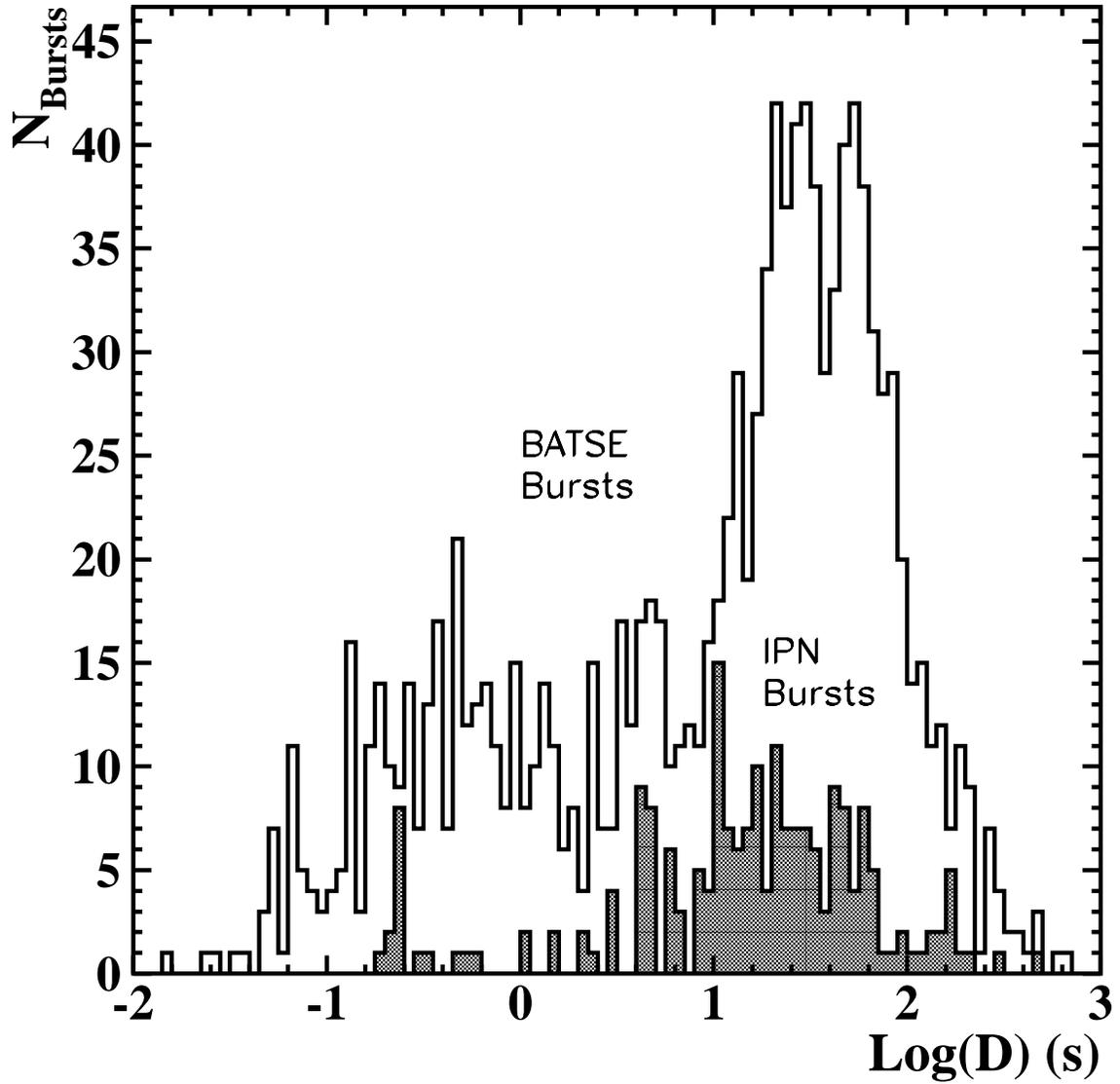}
\caption{Duration distribution of BATSE GRBs (upper, open histogram) and 
IPN bursts for which durations have been determined (lower, shaded 
histogram).  Both distributions appear to be drawn from the same 
underlying population.
\label{fig8}}
\end{figure}

\begin{figure}
\epsscale{0.95}
\plotone{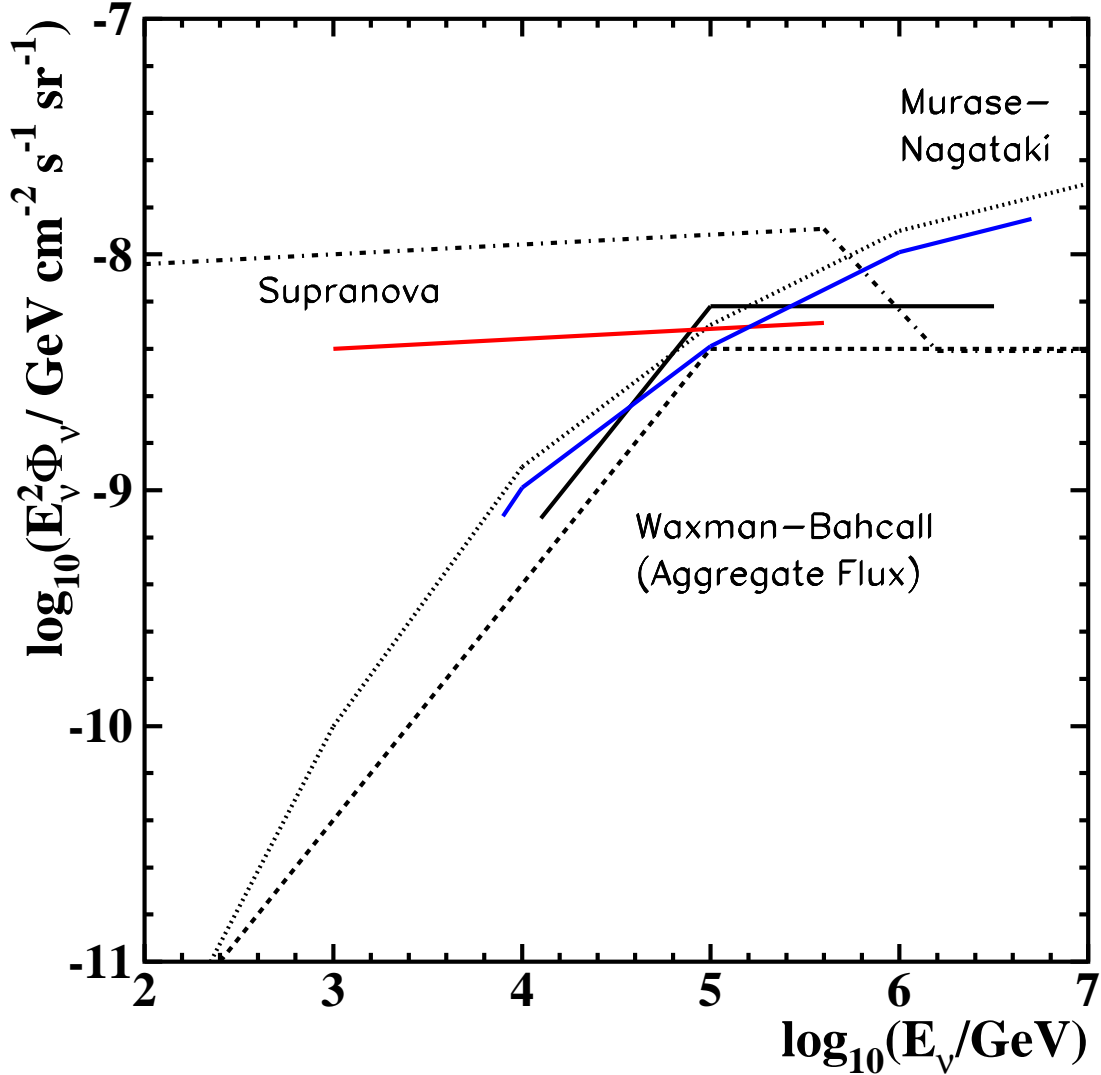}
\caption{AMANDA flux upper limits (solid lines) for muon neutrino energy spectra 
predicted by the Waxman-Bahcall spectrum \citep{Waxman} (thick dashed line), the Razzaque 
et al. spectrum \citep{Razzaq} (dot-dashed line) and the Murase-Nagataki spectrum 
\citep{Murase} (thin dotted line).  The central 90\% of the expected flux for each model is 
shown.  For the Waxman-Bahcall model we include both long- and short-duration bursts; for 
the other spectra, only long-duration bursts are included.  Including short-duration bursts 
would improve the flux upper limits by approximately 13\%.  While our analysis was 
restricted to bursts located in the Northern Hemisphere (2${\pi}$ sr), all flux upper 
limits are for the entire sky (4${\pi}$ sr).  
\label{fig9}}
\end{figure}

\begin{figure}
\epsscale{1.0}
\plotone{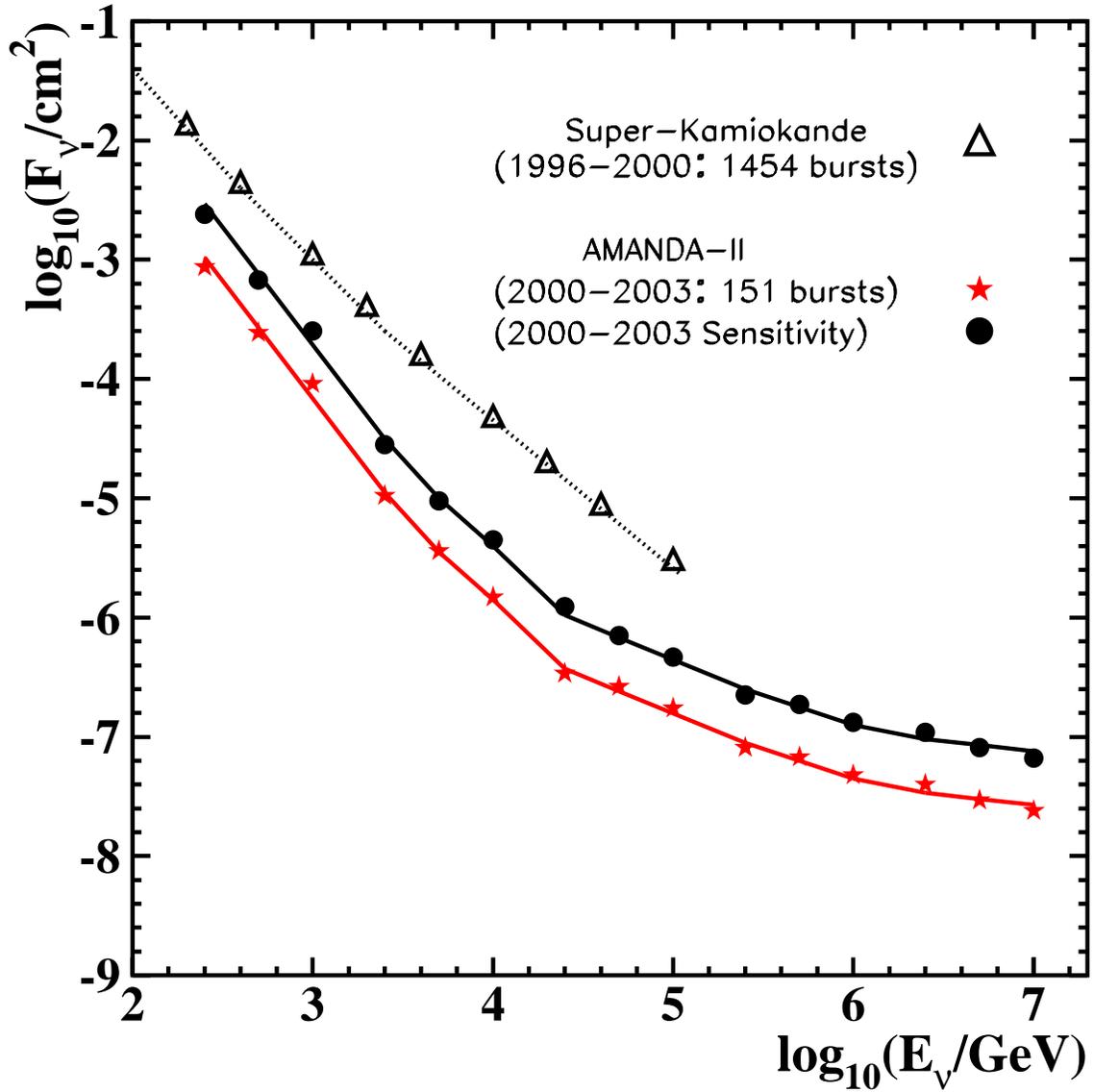}
\caption{Green's Function Fluence Upper Limit for AMANDA's GRB analysis from 2000 to 2003.  
This fluence upper limit can be folded into any desired spectrum to provide a flux upper 
limit for that particular spectrum.
\label{fig10}}
\end{figure}

\clearpage

\begin{table}
\begin{center}
\caption{Primary Instruments in the Third Interplanetary Network, 1997-2003 \label{tab:a}}
\begin{tabular}{ccc}
\tableline
Instrument & Energy Range (keV) & Mission Homepage \\
\hline
BATSE LAD & 30 - 190 & http://www.batse.msfc.nasa.gov \\
BeppoSAX GRBM & 40 - 700 & http://www.asdc.asi.it/bepposax \\
BeppoSAX WFC & 2 - 26 & http://www.asdc.asi.it/bepposax \\
HETE-II FREGATE & 6 - 400 & http://space.mit.edu/HETE/fregate.html \\
HETE-II WXM & 2 - 25 & http://space.mit.edu/HETE/wxm.html \\
HETE-II SXC & 2 - 14 & http://space.mit.edu/HETE/sxc.html \\
INTEGRAL & 15 - 10000 & http://integral.esac.int/ \\
Konus WIND & 12 - 10000 & http://www-spof.gsfc.nasa.gov/istp/wind/ \\
Mars Odyssey & $\sim$100 - 8000 & http://mars.jpl.nasa.gov/odyssey/ \\
NEAR XGRS & 100 - 1000 & http://near.jhuapl.edu \\
RHESSI & $\sim$25 - $\sim$25000 & http://hesperia.gsfc.nasa.gov/hessi \\
Ulysses & 25 - 150 & http://ulysses.jpl.nasa.gov \\
\hline
\end{tabular}
\end{center}
\end{table}

\clearpage

\begin{table}
\begin{center}
\caption{BATSE Triggered and IPN Bursts Per Year in the AMANDA Analysis, by Duration \label{tab:b}}
\begin{tabular}{cccccccc}
\tableline
Year & 1997 & 1998 & 1999 & 2000 & 2001 & 2002 & 2003 \\
\hline
N$_{\mathrm{Short}}$ & 12 & 15 & 9 & 7 & 1 & 1 & 2 \\
N$_{\mathrm{Long}}$ & 51 & 50 & 61 & 77 & 15 & 21 & 24 \\
N$_{\mathrm{Unknown}}$ & 15 & 29 & 26 & 3 & 0 & 0 & 0 \\
\hline
N$_{\mathrm{Total}}$ & 78 & 94 & 96 & 87 & 16 & 22 & 26 \\
\hline
\end{tabular}
\end{center}
\end{table}

\clearpage

\begin{table}
\begin{center}
\caption{Data Selection Criteria, Year by Year \label{tab:c}}
\begin{tabular}{cccccc}
\tableline
Criterion & 97-99 & 00 & 01-03 & Precursor\tablenotemark{a} \\
\hline
${\Delta}$${\Psi}_1$, ${\delta}$${\geq}$10$^\circ$ (${\delta}$$<$10$^\circ$) & $<$20$^\circ$ ($<$6.5$^\circ$) & 
$<$12.5$^\circ$ ($<$7$^\circ$) & $<$12$^\circ$ ($<$8$^\circ$) & $<$12$^\circ$ ($<$5$^\circ$) \\
${\Delta}$${\Psi}_2$, ${\delta}$${\geq}$10$^\circ$ (${\delta}$$<$10$^\circ$) & N/A & N/A & $<$12$^\circ$ ($<$8$^\circ$) & 
$<$12$^\circ$ ($<$6$^\circ$) \\
${\Delta}$${\Psi}_3$, ${\delta}$${\geq}$10$^\circ$ (${\delta}$$<$10$^\circ$) & N/A & N/A & $<$16$^\circ$ ($<$8$^\circ$) & 
$<$16$^\circ$ ($<$8$^\circ$) \\
${\Delta}$${\Psi}_4$, ${\delta}$${\geq}$10$^\circ$ (${\delta}$$<$10$^\circ$) & N/A & N/A & N/A & $<$40$^\circ$ 
($<$40$^\circ$) \\
${\sigma}_{\Psi}$\tablenotemark{b} & N/A & N/A & $<$5$^\circ$ ($<$5$^\circ$) & $<$5$^\circ$ ($<$5$^\circ$) \\
Track Uniformity & N/A & $<$0.29 ($<$0.29) & $<$0.55 ($<$0.55) & $<$0.55 ($<$0.55) \\
${\pounds}_{reco}$\tablenotemark{c} & N/A & $<$7.85 ($<$7.5) & N/A & N/A \\
Direct Hits & $>$10 & N/A & N/A & N/A \\
$N_{OMs}$ in Event & N/A & N/A ($>$24) & N/A & N/A \\
\hline
Signal Passing Rate & 0.35 (0.22)& 0.69 (0.54) & 0.68 (0.61) & 0.69 \\
\hline
\end{tabular}
\tablenotetext{a}{{The precursor time period was searched only during the 2001-2003 dataset.}}
\tablenotetext{b}{{The angular resolution of the paraboloid fit.}}
\tablenotetext{c}{{The log(Likelihood) of the reconstructed track.}}
\end{center}
\end{table}

\clearpage

\begin{table}
\begin{center}
\caption{Uncertainties/Corrections in the GRB Analysis \label{tab:d}}
\begin{tabular}{ccc}
\tableline
Source of Uncertainty & Quantity & Reference \\
\hline
OM sensitivity & ${\pm}$7\% & \citep{Reco} \\
Simulation parameters (incl. ice properties) & ${\pm}$15\% & Sections 3.3 \& 3.4 \\
Neutrino-nucleon cross-section & ${\pm}$3\% & \citep{Gandhi} \\
Correction for IPN bursts not modeled & -3\% & Appendix A \\
\hline
Uncertainties added in quadrature & +16/-17\% & \\
\hline
Correction for short bursts not modeled & -6\% & Appendix A \\
\hline
Total & +15\%/-18\% & \\
\hline
\end{tabular}
\end{center}
\end{table}

\clearpage

\begin{table}
\begin{center}
\caption{Results of GRB Analysis 1997-2003 \label{tab:e}}
\begin{tabular}{cccccc}
\tableline
Year & 1997-1999 & 2000 & 2001-2003 & 2000-2003 & 1997-2003 \\
\hline
$N_{\mathrm{Bursts}}$ & 268 & 87 & 64 & 151 & 419 \\
$N_{\mathrm{BG,Exp}}$ & 0.46 & 1.02 & 0.27 & 1.29 & 1.74 \\
$N_{\mathrm{Obs}}$ & 0 & 0 & 0 & 0 & 0 \\
Event Upper Limit & 1.98 & 1.50 & 2.30 & 1.30 & 1.10 \\
$MRF_{\mathrm{WB}}$\tablenotemark{a} & 6.6 & 5.5 & 11 & 2.5 & 1.3 \\
$MRF_{\mathrm{MN}}$\tablenotemark{b} & 4.9 & 3.1 & 6.2 & 1.4 & 0.82 \\
$MRF_{\mathrm{Razz}}$\tablenotemark{c} & 2.4 & 1.5 & 3.0 & 0.68 & 0.40 \\
\hline
\end{tabular}
\tablenotetext{a}{{Based on the flux of \citet{Waxman}, corrected for neutrino oscillations.}}
\tablenotetext{b}{{Based on the flux of \citet{Murase}.}}
\tablenotetext{c}{{Based on the ``supranova'' flux of \citet{Razzaq}.}}
\end{center}
\end{table}

\clearpage

\begin{table}
\begin{center}
\caption{Results of Precursor Search 2001-2003 \label{tab:f}}
\begin{tabular}{ccccc}
\tableline
Year & $N_{\mathrm{Bursts}}$ & $N_{\mathrm{BG,Exp}}$ & $N_{\mathrm{Obs}}$ & Event U.L. \\
\hline
2001 & 15 & 0.06 & 0 & 2.38 \\
2002 & 21 & 0.07 & 0 & 2.37 \\
2003 & 24 & 0.07 & 0 & 2.37 \\
\hline
2001-2003 & 60 & 0.20 & 0 & 2.30 \\
\hline
\end{tabular}
\end{center}
\end{table}

\end{document}